\documentclass[
  journal=pasa,
  manuscript=research-paper,
  year=2020,
  volume=37,
]{cup-journal}
\usepackage{subfig}
\usepackage{float}
\usepackage{amsmath,amssymb}
\usepackage[nopatch]{microtype}
\usepackage{booktabs}
\usepackage{multirow}
\usepackage{array}
\newcolumntype{P}[1]{>{\raggedright\arraybackslash}p{#1}}
\newcolumntype{A}[1]{>{\raggedleft\arraybackslash}p{#1}}
\usepackage{mathtools}
\usepackage{longtable}
\usepackage[thinc]{esdiff}

\newcommand{\bz}{$\langle B_z \rangle$}

\newcommand\farcm{\hbox{$.\!^{\prime}$}}
\newcommand\farcs{\hbox{$.\!\!^{\prime\prime}$}}
\def\arcsec{\hbox{$^{\prime\prime}$}}

\usepackage{hyperref}
\hypersetup{colorlinks,citecolor=blue,linkcolor=blue,urlcolor=blue}

\title{VAST-MeMeS: Characterising non-thermal radio emission from magnetic massive stars using the Australian SKA Pathfinder}

\author{Barnali Das}
\affiliation{CSIRO, Space and Astronomy, P.O. Box 1130, Bentley WA 6102, Australia}
\email[Barnali Das]{Barnali.Das@csiro.au}

\author{Laura~N. Driessen}
\affiliation{Sydney Institute for Astronomy, School of Physics, The University of Sydney, New South Wales 2006, Australia}

\author{Matt E. Shultz}
\affiliation{Department of Physics and Astronomy, University of Delaware, 217 Sharp Lab, Newark, Delaware, 19716, USA}

\author{Joshua Pritchard}
\affiliation{Australian Telescope National Facility, CSIRO, Space and Astronomy, PO Box 76, Epping, 1710, NSW, Australia}

\author{Kovi Rose}
\affiliation{Sydney Institute for Astronomy, School of Physics, The University of Sydney, New South Wales 2006, Australia}
\alsoaffiliation{Australian Telescope National Facility, CSIRO, Space and Astronomy, PO Box 76, Epping, 1710, NSW, Australia}

\author{Yuanming Wang}
\affiliation{Centre for Astrophysics and Supercomputing, Swinburne University of
Technology, Hawthorn, Victoria, 3122, Australia}

\author{Yu Wing Joshua Lee}
\affiliation{Sydney Institute for Astronomy, School of Physics, The University of Sydney, New South Wales 2006, Australia}
\alsoaffiliation{ARC Centre of Excellence for Gravitational Wave Discovery (OzGrav), Hawthorn, 3122, Victoria, Australia}
\alsoaffiliation{Australian Telescope National Facility, CSIRO, Space and Astronomy, PO Box 76, Epping, 1710, NSW, Australia}

\author{Gregory Sivakoff}
\affiliation{Department of Physics, University of Alberta, CCIS 4-181, Edmonton AB T6G 2E1, Canada}

\author{Andrew Zic}
\affiliation{Australian Telescope National Facility, CSIRO, Space and Astronomy, PO Box 76, Epping, 1710, NSW, Australia}

\author{Tara Murphy}
\affiliation{Sydney Institute for Astronomy, School of Physics, The University of Sydney, New South Wales 2006, Australia}

\keywords{stars: early-type stars, stars: magnetic field, radio continuum: stars, stars: massive, stars: variable} 

\begin{document}

\begin{abstract}
Magnetic massive stars are stars of spectral types O, B and A that harbour $\sim$ kG strength (mostly dipolar) surface magnetic fields.
Their non-thermal radio emission has been demonstrated to be an important magnetospheric probe, provided the emission is fully characterised. A necessary step for that is to build a statistically significant sample of radio-bright magnetic massive stars.
In this paper, we present the `VAST project to study Magnetic Massive Stars' or VAST-MeMeS that aims to achieve that by taking advantage of survey data acquired with the Australian SKA Pathfinder telescope. VAST-MeMeS is defined under the `Variables and Slow Transients' (VAST) survey, although it also uses data from other ASKAP surveys.
We found radio detections from 48 magnetic massive stars, out of which, 14 do not have any prior radio detections. We also identified 9 `Main-sequence Radio Pulse Emitter' candidates based on variability and circular polarisation of flux densities.
The expanded sample suggests a slightly lower efficiency in the radio production than that reported in earlier work.
In addition to significantly expanding the sample of radio-bright magnetic massive stars, the addition of flux density measurements at $\lesssim 1$ GHz revealed that the spectra of incoherent radio emission can extend to much lower frequencies than that assumed in the past. 
In the future, radio observations spanning wide frequency and rotational phase ranges should be conducted so as to reduce the uncertainties in the incoherent radio luminosities. The results from these campaigns, supplemented with precise estimations of stellar parameters, will allow us to fully understand particle acceleration and non-thermal radio production in large-scale stellar magnetospheres.
\end{abstract}


\section{Introduction}\label{sec:intro}
Large-scale, ordered magnetic fields are found in approximately 10\% of the early-type (spectral type OBA) star population \citep{grunhut2012}. Unlike the magnetic fields observed in solar-type stars, these fields are extremely stable in time \citep[e.g.][]{braithwaite2004,wade2016,shultz2018}. The radiatively driven stellar wind of OBA stars interacts with the strong ($\sim \mathrm{kG}$) magnetic fields, leading to the formation of co-rotating magnetospheres. 

\citet{petit2013} introduced a classification scheme for these magnetospheres based on the relative values of two parameters: the Kepler radius $R_\mathrm{K}$, where centrifugal force due to co-rotation balances gravity, and the Alfv\'en radius $R_\mathrm{A}$ that defines the extent of the largest closed magnetic field line. 
For stars with $R_\mathrm{A}<R_\mathrm{K}$, the stellar wind materials ejected along the closed field loops follow the magnetic field lines and then fall back to the star in dynamical timescale. This type of magnetosphere is named as `dynamical magnetosphere' (DM). On the other hand, if $R_\mathrm{A}>R_\mathrm{K}$, there will be a region inside the magnetosphere, defined by $R_\mathrm{K}<r<R_\mathrm{A}$ ($r$ is the radial distance from stellar centre along the magnetic equator), where the outward centrifugal force is stronger than the inward gravity. Such a region is named as `centrifugal magnetosphere' (CM).
Inside the CM, the continuous injection of mass flow due to stellar wind is balanced by small-scale explosions called centrifugal breakout (CBO), during which the excess plasma breaks open the magnetic field lines temporarily, and escapes the star \citep{shultz2020,owocki2020}.

CBOs are small spatial-scale events that are present at all times. As such, we do not expect to see explosions directly as flares unless we resolve the stellar magnetosphere. Their presence was indirectly inferred based on H$\alpha$ observations \citep{shultz2020,owocki2020}. However, magnetic late B- and A-type stars do not produce detectable H$\alpha$ emission, raising the question of whether a similar operating scenario applies to their magnetospheres.
Fortunately, magnetospheric emission occurs over a wide spectral band including radio, and radio emission has been observed even from much cooler magnetic early-type stars.
Only recently, it was shown that magnetic reconnection triggered by CBOs can explain the incoherent radio luminosities of all magnetic hot stars, establishing CBOs to be a ubiquitous magnetospheric phenomenon across the OBA spectral types \citep{leto2021,shultz2022,owocki2022}. This was supported by the observation that all solitary, radio-bright hot magnetic stars harbour CMs \citep{shultz2022}.

The new scenario of CBO-driven radio production critically relies on an empirical relation between incoherent radio luminosity and a combination of stellar parameters: magnetic field, stellar radius, and rotation period. No evidence for the role of temperature or the mass-loss rate was obtained, which was non-intuitive since the electrons emitting the radio emission are provided by the stellar wind \citep{leto2021,shultz2022}. The CBO theory can explain this trend only under the assumption that the magnetic field topology behaves like a monopole at the reconnection sites \citep{owocki2022}. At the moment, it is unclear what might lead to such a situation. The empirical relation also has a large scatter, suggesting involvement of additional stellar parameters \citep{shultz2022,keszthelyi2024}.

In addition to incoherent emission, a subset of the magnetic hot stars, called `Main-sequence Radio Pulse emitters' \citep[MRPs,][]{das2021}, also emit periodic coherent radio pulses, produced by electron cyclotron maser emission. All MRPs are found to also produce incoherent radio emission. However, the coherent emission appears to have a dependence on temperature \citep{das2022}, unlike incoherent emission. In particular, coherent emission is less efficient in stars with $T_\mathrm{eff}\gtrsim 19$ kK \citep{das2022c}. Note that the sample size of MRPs is only 19, and thus a robust conclusion can only be drawn after expanding the sample.

Similar to the case of coherent emission, the empirical relation derived for incoherent emission also needs to be tested with a larger sample considering its profound implications for magnetospheric physics. The largest sample size used to examine the correlation considered only the 48 detected stars \citep{keszthelyi2024,shultz2022}. Interestingly, even after considering non-detections, the number of hot magnetic stars \textit{investigated} at radio bands constitute $<20\%$ of the known hot magnetic star population \citep[][Shultz et al. in prep.]{shultz2022}.

In this paper, we introduce the `VAST project to study Magnetic Massive Stars' or VAST-MeMeS project that aims to solve this key issue by detecting new radio-bright magnetic massive stars, MRPs, and also by providing useful constraints on the radio luminosities of hot magnetic stars by using the massive amount of data being provided by the new-generation Australian SKA Pathfinder \citep[ASKAP\footnote{\href{https://www.atnf.csiro.au/projects/askap/index.html}{https://www.atnf.csiro.au/projects/askap/index.html}};][]{2021PASA...38....9H} telescope. This paper is structured as follows. In the next section (\S\ref{sec:vast-memes}), we explain this project in greater detail. Section \ref{sec:obs} focuses on the data used in this paper. This is followed by results (\S\ref{sec:results}) and discussion (\S\ref{sec:discussion}). We summarise the paper in \S\ref{sec:summary}.

\section{VAST-MeMeS}\label{sec:vast-memes}
ASKAP is a 36-dish interferometer where each dish is 12 metre in diameter. It is a survey instrument with a field of view of $\sim30$\,square\,degrees and a bandwidth of 288\,MHz.
VAST-MeMeS is a project defined under the ASKAP Variables And Slow Transients (VAST) survey \citep{murphy2013}. VAST is dedicated to untargeted searches for transients/variables at timescales $\sim \mathrm{minutes} $ to years. The survey's strategy is to observe chosen patches of the sky at multiple epochs over the frequency range of 743.5--1031.5 MHz (central frequency of 887.5 MHz), each such epoch has a duration of 12 minutes. 
This is significantly smaller than the rotation period of the fastest magnetic hot stars known \citep[$\approx 12$ hours, e.g.][]{grunhut2012b}.
Thus, the radio continuum images from different epochs can be compared to detect variable sources (rotational modulation).

We used data that are publicly available in the ASKAP data archive known as CSIRO ASKAP Science Data Archive (CASDA\footnote{\url{https://research.csiro.au/casda/}}). These data include observations with a range of integration times and central frequencies. Some observations are a few minutes while some are up to 12 hours. The most common integration time is $\sim15$ minutes due to VAST and the Rapid ASKAP Continuum Survey \citep[RACS;][]{mcconnell2020}. The ASKAP frequency band is 288 MHz wide and the central frequency can range of $\sim800$\,MHz to $\sim1700$\,MHz.

At our observing frequencies, radio emission from magnetic hot stars is expected to be entirely non-thermal in nature, except for massive O-stars.
The overall frequency range of data provided by ASKAP offers another advantage. \citet{leto2021} showed that the incoherent radio spectra of hot magnetic stars tend to exhibit a turnover at $\approx 1$ GHz (see their Figure 2). On the other hand, coherent radio emission has been found to be much more prevalent at frequencies $\lesssim 1$ GHz \citep[e.g.][]{das2020b,das2021} and the emission seems to be much weaker above 3 GHz \citep{das2022b}. Thus, our frequency range is suitable for discovering both coherent and incoherent radio emission.

An important part of the VAST-MeMeS project is a recently made catalogue of known magnetic hot stars (Shultz et al. in prep.) that includes $761$ OBA stars with at least one magnetic detection. This catalogue also provides information on the stellar magnetic field, rotation period, bolometric luminosity, surface temperature, surface gravity, inclination angles and obliquity for stars with reported values. The stellar radii are calculated using bolometric luminosity and effective temperature, which is then used to calculate the stellar mass from surface gravity.

The specific goals of this project are to use the enormous data-set that ASKAP will provide in the coming years for the following purposes:
\begin{enumerate}
    \item Discover new radio-bright magnetic massive stars;
    \item Discover new MRPs (coherent radio emission from magnetic hot stars);
    \item Explore the possibility of transients (flares) from magnetic massive stars;
    \item Discover new magnetic massive star candidates.    
\end{enumerate}

The observational signatures of different types of radio emission produced by magnetic hot stars, which we will use to achieve our goals, are described in the following subsections.

\subsection{Incoherent radio emission}\label{subsubsec:incoherent}
This is the most common type of radio emission at our observing frequencies for magnetic hot stars. It is expected to be weakly circularly polarised at $\sim 1$ GHz \citep[$\lesssim 10\%$, e.g.][]{leto2017,leto2018}. Both total intensity and circular polarisation exhibit rotational modulation and the strength of that modulation decreases with decreasing frequencies \citep[e.g.][]{leto2017,leto2018}. 
The modulation also correlates with that of the stellar longitudinal magnetic field \bz~in the sense that the flux densities are maximum at (or close to) the extrema of \bz~and minimum when \bz~is closest to zero \citep[e.g.][]{lim1996}.

\subsection{Coherent radio emission}\label{subsubsec:coherent}
Coherent radio emission is typically observed as highly circularly polarised pulses. Thus, they have two properties that set them apart from incoherent emission: flux density variation over a much shorter time-scale and high circular polarisation. In addition, the flux density variation of coherent emission is more prominent than that of incoherent emission, especially at our observing frequencies.

Under VAST-MeMeS, we employ the following criteria to identify MRP candidates:
\begin{enumerate}
    \item High circular polarisation ($>30\%$, see \S\ref{sec: circ pol search})
    \item For stars with multi-epoch observations, if the flux density changes by a factor $>2$ between different epochs for the same observing frequency.
    \item For stars with multi-epoch observations and with available rotation period measurements, if the flux density changes by a factor $\geq 1.5$ within a rotational phase window of width $\leq 0.16$ for the same observing frequency.
\end{enumerate}
The first criterion is based on the case of HD\,37017 for which coherent pulses have been observed with fractional circular polarisation as low as $37\pm 11 \%$ \citep{das2022}. The second criterion is influenced from the observation of HD\,182180, a known MRP, which exhibits clear rotational modulation in its incoherent radio emission even at $0.7$ GHz, with a modulation factor (ratio of maximum to minimum flux density) of $\approx 1.7$ \citep{das2022c}. The third criterion is based on the `minimum flux-density gradient condition' introduced by \citet{das2022}: $\Delta\phi_\mathrm{rot}<0.16$, where $\Delta\phi_\mathrm{rot}$ is the rotational phase range over which the pulse rises from its basal to peak flux density.
The second and the third criteria are somewhat lenient in the sense that they do not rule out incoherent emission. However, this will minimise the probability of missing any MRPs in the sample. 

We consider a magnetic hot star as an MRP candidate if it satisfies any of the above three criteria.



\subsection{Radio flares from magnetic massive stars}\label{subsubsec:flares}
For a long time, magnetic massive stars have been considered to be objects that do not exhibit flares, i.e. enhancements in flux densities that are not predictable (unlike the radio pulses described in the preceding section, which are periodic). However, with the introduction of the scenario of centrifugal breakout as the primary mechanism for plasma transport for radio-bright magnetic massive stars \citep{shultz2022,owocki2022}, the possibility of flares from magnetic massive stars cannot be summarily ruled out. \citet{das2021} reported observation of highly circularly polarised radio flares from the \textit{direction} of the magnetic hot star CU\,Vir for the first time. This was followed by reports of radio flares from another three magnetic hot stars by \citet{polisensky2023}. 
The unique observation strategy of the VAST survey (multi-epoch observations) will also allow us to look for flares from magnetic massive stars at radio bands. Note that the MRP candidates that will be obtained following the procedure outlined in the preceding subsection will also be candidates for flaring magnetic massive stars. The nature of the flux density enhancements, i.e. whether or not they are periodic with the stellar rotation period, can be confirmed by follow-up observations.

\subsection{Discovery of new magnetic massive stars via radio emission}\label{subsec:new_mms}
Non-thermal radio emission is a key indicator of magnetism, and this aspect can be used to discover new magnetic massive star candidates. For binary systems involving two massive stars, the collision between their strong winds can give rise to synchrotron radio emission. O-type stars are also known to produce free-free radio emission owing to their high mass-loss rates. But such emission is expected to be less prominent at sub-GHz frequencies.

Under VAST-MeMeS, we will examine the following properties, where available, to discover potentially magnetic massive stars:
\begin{enumerate}
    \item Temporal variation of radio emission;
    \item Modulation with rotational phase;
    \item Percentage circular polarisation.
\end{enumerate}
Based on the above properties, we will identify the most-probable magnetic star candidates, which can then be followed up with spectro-polarimetric observations. 

In this first paper of VAST-MeMeS, we focus on incoherent and coherent radio emission from known magnetic hot stars.

\section{Data analysis}\label{sec:obs}

We used two methods to identify radio emission from the $761$ hot magnetic stars identified by Shultz et al as well as the 9 other known radio emitting hot magnetic stars. We searched for circularly polarised (Stokes V) radio emission at the positions of the stars, and we cross-matched the positions of the stars to ASKAP Stokes I point sources.

\subsection{Cross-matching}
\label{sec: cross-matching}

In order to cross-match to identify radio emission from hot magnetic stars, we first need to create a catalogue of ASKAP radio sources and perform a Monte Carlo simulation to determine the appropriate cross-match radius.

We used all of the \texttt{Selavy}\footnote{\url{https://www.atnf.csiro.au/computing/software/askapsoft/sdp/docs/current/analysis/selavy.html}} \citep{2012MNRAS.421.3242W,2012PASA...29..371W} source catalogues available as of 2024 October 30 in the CASDA. We included all publicly available data where the quality was either "good" or "uncertain". 

We combined all of the source catalogues into three catalogues:
\begin{itemize}
    \item One catalogue including every source detected (total catalogue);
    \item One catalogue including every source with $S_{int} / S_{peak} <= 1.5$, \textsc{has\_siblings=0} and the integration time of the observation was $\leq1000$ sec (short catalogue);
    \item One catalogue including every source with $S_{int} / S_{peak} <= 1.5$, \textsc{has\_siblings=0} and the integration time of the observation was $>1000$ sec (long catalogue).
\end{itemize}
where $S_{int}$ and $S_{peak}$ are the integrated and peak flux densities respectively. If a source has $S_{int} / S_{peak} <= 1.5$, it is more likely to be unresolved. The \textsc{has\_siblings} flag is calculated in \texttt{Selavy}. Sources with \textsc{has\_siblings}=0 are more likely to be unresolved as they do not have nearby associated sources. We split the unresolved source catalogues by integration time as the density of sources varies with integration time. ASKAP has observed many fields with an integration time of either $\sim15$ minutes or $\gtrsim10$ hours. This is why we divided the integration time into these two bins.

However, many of the ASKAP observations are repeat observations of the same fields. This means for each radio source there may be multiple detections. In order to calculate the reliability radii for cross-matching, we required catalogues with each source included only once. As we are looking for detections of stars, we also expected the radio detections to be unresolved (point sources). Some ASKAP fields have long integration time images $\gtrsim10$ hours while others have only been observed on short time scales of $\lesssim15$ minutes. If a point source in a 15 minute observation appears as an extended source in a 10 hour observation, we wanted to exclude that source from our point source catalogues. On the other hand, a source in a short observation close to the signal to noise limit of five may not meet our $S_{int} / S_{peak} <= 1.5$ requirement in the short observation but may meet it in a longer integration observation. This made it challenging to determine which sources to include as point sources and which should be excluded as extended sources. Selecting the best position for a source was also challenging as there were systematic shifts in position between observations.

To determine if a source was resolved or unresolved, we matched every source in the long and short catalogues to sources in the total catalogue. We considered a source to be the same source if the separation was less than $\frac{a_1 + a_2}{2}$ where $a_1$ was the semi-major axis of the source in the long or short catalogue and $a_2$ was the semi-major axis of the source in the total catalogue. Each long/short point source may have multiple total catalogue sources associated with it. If the associated source with the highest signal to noise met the $S_{int} / S_{peak} <= 1.5$ requirement we kept the corresponding short/long source. If not, the source was removed from the long/short catalogue. If the source was kept, we calculated the median position of the associated sources and used that as the best position of the source for the purposes of the Monte Carlo to determine the cross-match radius. We removed duplicate sources from the short/long catalogues if two sources were separated by less than $\frac{a_1 + a_2}{2}$ as above. After this process, we were left with two catalogues of unresolved sources, the short and long integration catalogues. These were the catalogues used to calculate the appropriate cross-match radii. There were 3,690,531 sources in the short catalogue and 5,555,927 sources in the long catalogue.

\subsubsection{Cross-match reliability}
\label{sec: MC}

We now have two radio point source catalogues and the positions of the hot magnetic stars. We performed a Monte Carlo simulation using the median positions of the radio sources to determine the cross-match radius that results in a reliability of 98\%.

\begin{figure}
    \centering
    \includegraphics[width=\columnwidth]{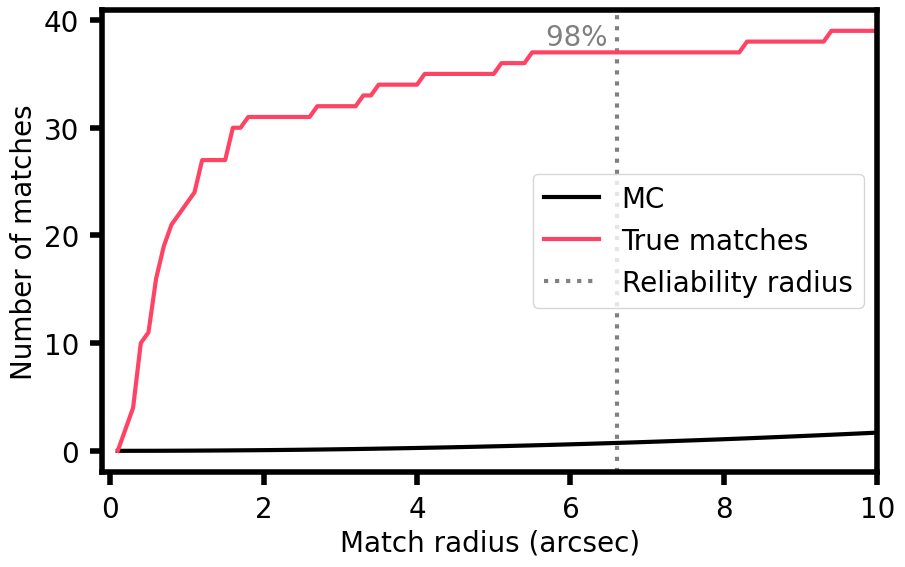}
    \caption{Cumulative cross-match results for the short catalogue cross-match to the positions of the known hot magnetic stars. The black line shows the results of the 100,000 iteration Monte Carlo simulation. The red line shows the results of the cross-matches when the true coordinates of the radio sources and hot magnetic stars are used. The radio positions used for this are the median positions and an epoch of J2022 was used to correct for proper motion. The grey-dashed line shows the radius where the reliability is 98\%.}
    \label{fig:short cat MC}
\end{figure}

\begin{figure}
    \centering
    \includegraphics[width=\columnwidth]{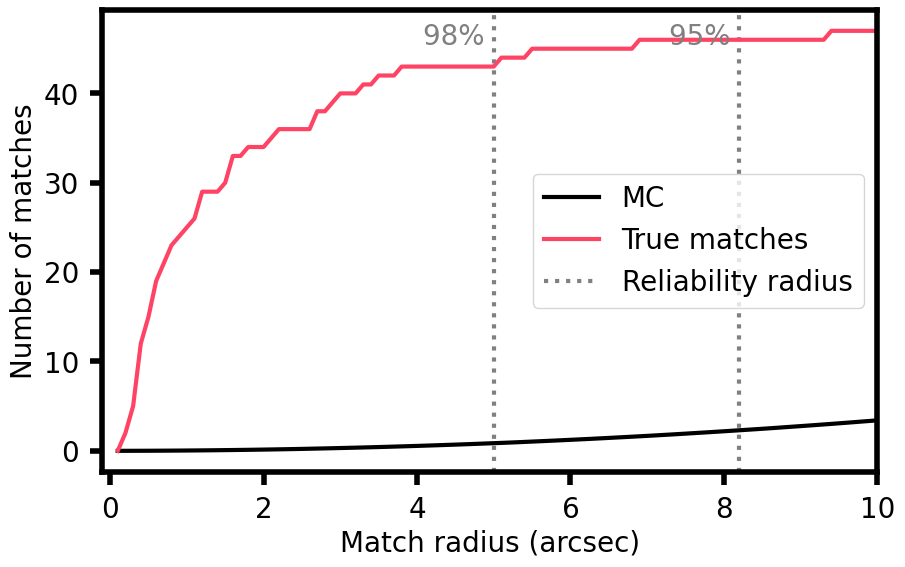}
    \caption{Cumulative cross-match results for the long catalogue cross-match to the positions of the known hot magnetic stars. The black line shows the results of the 100,000 iteration Monte Carlo simulation. The red line shows the results of the cross-matches when the true coordinates of the radio sources and hot magnetic stars are used. The radio positions used for this are the median positions and an epoch of J2022 was used to correct for proper motion. The grey-dashed line shows the radius where the reliability is 98\%.}
    \label{fig:long cat MC}
\end{figure}

We used the same method presented in \citet{driessen2024} treating the long and short catalogues independently. We iterated over the method 100,000 times and used a minimum and maximum shift radius of 0\farcm5 and 1\farcm5. To determine the ``true matches'' for this method we used the median positions of the point sources and an epoch of J2022 to account for proper motion. 
The reliability was calculated using $R=1-N_{\mathrm{random,r}}/N_{\mathrm{true,r}}$, where $R$ is the reliability, $N_{\mathrm{random,r}}$ is the number of matches at a given radius, $r$, from the 100,000 iteration Monte Carlo simulation, and $N_{\mathrm{true,r}}$ is the number of matches $r$ using the true hot magnetic star and median radio source positions and a proper motion correction epoch of J2022.
The cross-match radius required for a reliability of 98\% was 6\farcs6 for the short ASKAP catalogue and 5\farcs0 for the long ASKAP catalogue. The results of the short and long catalogue Monte Carlo simulations are shown in Figures\,\ref{fig:short cat MC} and \ref{fig:long cat MC} respectively. The typical ASKAP systematic position uncertainties are 1-2\arcsec. As both of these 98\% reliability radii are larger than 2\arcsec, we can use these as our cross-match radii to match the short and long catalogue source positions to the hot magnetic star positions.

\subsubsection{Cross-matching results}
\label{sec: cross-match results}

We used the 98\% reliability radii to perform the final cross-matching. We used the original positions and detection epochs of the radio sources to account for proper motions, rather than using the median positions for each unique radio source. This means that we use the short and long catalogues prior to removing the duplicate detections. It also means that the 98\% cross-match radii are likely a slight underestimate. The number of ``true'' matches used to calculate the reliability relied on the median position being an approximation of the source position at the J2022 epoch. However, this may mean ``true'' matches were missed if the star's proper motion was larger.

We identified radio emission from 39 stars using the short catalogue and 22 stars using the long catalogue. This results in a total of 48 unique stars. The stars identified using the two radio catalogues are shown in Tables\,\ref{tab: short star results} and \ref{tab: long star results}.

\subsection{Circular polarisation search}
\label{sec: circ pol search}

Highly circularly polarised radio emission has been detected from hot stars so we used this as a search method to identify hot stars in ASKAP observations. We used the \textsc{RadioFluxTools}\footnote{\href{https://gitlab.com/Sunmish/radiofluxtools}{\url{https://gitlab.com/Sunmish/radiofluxtools}}} package to perform forced fitting on all available ASKAP Stokes V images. We did this by taking the position of every source in a \textsc{Selavy} catalogue with $S_{int} / S_{peak} <= 1.5$. We then extracted the flux density at that position in the Stokes V image corresponding to the Stokes I image the source was detected in. We performed this Stokes V source extraction for the greater than 11,000 Stokes V images publicly available in CASDA as of 2024 October 30.

We then cross-matched the hot star positions to radio sources with Stokes V detections. We only included sources where the Stokes I detection had $S_{int} / S_{peak} <= 1.5$ and \texttt{has\_siblings} = 0 as for the cross-matching in Section\,\ref{sec: cross-matching}. We also only included radio sources where $\frac{|V|}{I} >$15\%, where $V$ is the Stokes V flux density and $I$ is the Stokes I flux density, and the signal-to-noise of the Stokes V detection was $>5$. These criteria reduce the possibility of spurious detections. In addition, coherent radio emission from magnetic hot stars is typically highly circularly polarised.
We used the same cross-match radii determined in Section\,\ref{sec: MC}: 6\farcs6 for observations $<1000$ sec and 5\farcs0 for observations $>1000$ sec long. These are conservative cross-match radii as the chance coincidence with Stokes V sources is significantly lower due to the low sky density of circularly polarised radio sources.

We identified nine stars using this method.
The nine stars are shown in Table\,\ref{tab: stokes V star results}.

\subsection{Multi-epoch detections}
\label{sec: multi-epoch detections}

We combined the previously known radio-bright magnetic hot stars (hereafter referred as `radio stars') with the new radio stars identified in Sections\,\ref{sec: cross-match results} and \ref{sec: circ pol search}. This resulted in a total of 68 radio stars. However, four of the previously known radio stars are located above +50$^\circ$ Declination and are therefore outside ASKAP's observing area. We used the \textsc{RadioFluxTools} package to search for repeat Stokes I and V ASKAP detections of the remaining 64 known radio stars. We used the same package to perform forced fitting in Stokes I and V images where the sources were observed but not detected.

We searched for repeat detections by cross-matching sources in all of the available Stokes I \textsc{Selavy} source catalogues in CASDA to the positions of the 64 radio stars. We use a cross-match radius of 6\arcsec. As these stars are known radio stars, we do not apply the same filtering that was required in Section\,\ref{sec: cross-matching}. We do not remove any sources from the \textsc{Selavy} source catalogues prior to cross-matching. We found a total of 310 detections of 49 stars using this method.

We then performed forced fitting for all ASKAP observations of the radio-detected stars where the star was observed, but not detected. We performed this forced fitting using \textsc{RadioFluxTools} at the optical position of each star.
We could perform forced fitting where both the Stokes I image and noise image were available in CASDA. This means some epochs are missed where the noise image was not available.
This means that we now have all available ASKAP detections and non-detections in both Stokes I and V for the hot magnetic stars detected in the radio.

Our strategy to find radio-bright magnetic hot stars uses slightly different criteria from those of \citet{driessen2024} while preparing the catalogues. Our final catalogue includes all the radio-bright magnetic hot stars reported by \citet{driessen2024} except for HD\,148937, which resides in a complex environment. After visually examining the relevant images, this star was added to our catalogue.

\section{Results}\label{sec:results}
\begin{figure*}
    \centering
    \includegraphics[width=\textwidth]{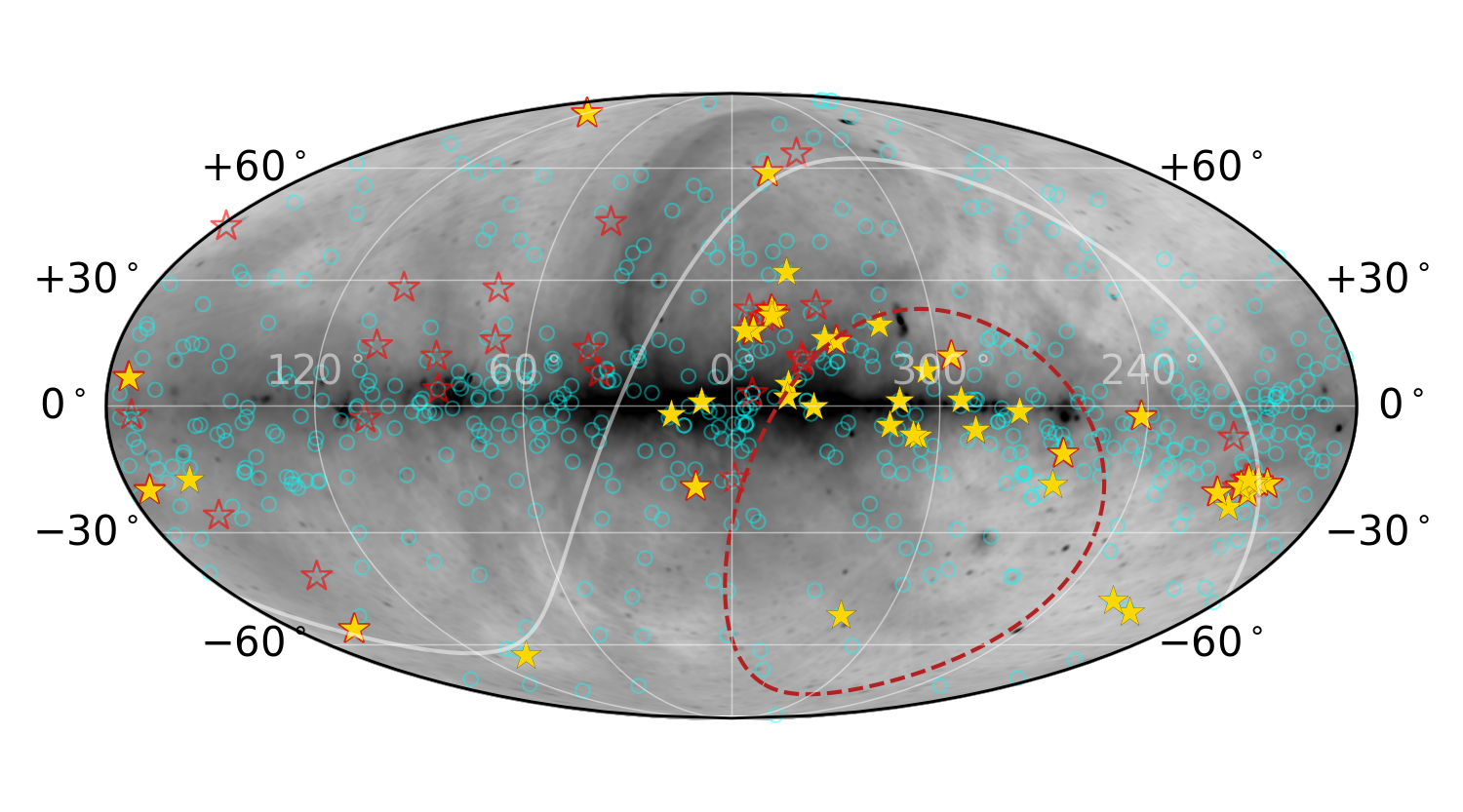}    
    \caption{The sky distribution (in galactic coordinates) of known magnetic massive stars (cyan unfilled circles, Shultz et al. in prep.), that of the radio-bright stars included in the sample of \citet{shultz2022} (shown in stars with thick red edges), and the ones detected by ASKAP (yellow-filled stars), including those reported by \citet{driessen2024}. The yellow stars with thick red edges represent the ones that are common to the sample of \citet{shultz2022} and the ASKAP sample. The red dashed contour represents the approximate declination limit for the Karl G. Jansky Very Large Array (VLA). The region inside the contour is inaccessible to the VLA.}
    \label{fig:askap_stars}
\end{figure*}

From the data acquired between 2018--02--17 and 2024--12--26 with the ASKAP, we obtained 
1109 measurements with 49 unique detections of magnetic massive stars, over frequencies ranging from 843 MHz to 1656 MHz. Among them, 25 stars were already included in the radio study of \citet{shultz2022}, an additional 10 were reported by \citet{biswas2023} (one star), \citet{driessen2024} (7 stars) and \citet{ayanabha2024} (2 stars).
The remaining 14 stars do not have any prior radio detections.
In addition, for 24 stars, the ASKAP detections mark the first radio detections within our observed frequency range ($\sim 1$ GHz). 
With these new additions, the total number of radio-bright magnetic hot stars increases to 70. This includes the recently reported radio detection of HD\,55522 by \citet{keszthelyi2025} at 650 MHz.
Among them, 11 stars have declinations less than $-40^\circ$ \citep[previously, it was 3,][]{shultz2022}.
The sky distribution of the detected stars is shown in Figure \ref{fig:askap_stars}. 
This figure demonstrates how ASKAP has provided us with a new discovery space in the form of access to the whole Southern sky.

Figure \ref{fig:parameter_span} shows the distribution of the expanded sample of radio-bright magnetic hot stars along different parameters that have been proposed to be relevant for incoherent radio emission \citep[][]{leto2021,shultz2022}. Note that the stars for which a given stellar parameter is not available, are excluded from respective histograms.
In the current distribution, the median values of the magnetic field strength, stellar radius, rotation period and effective temperatures are approximately 4 kG, $ 2.5\,R_\odot$, $ 1.4$ days and $ 13.8$ kK respectively.

\subsection{Incoherent emission}\label{subsec:result_incoherent}
According to \citet{shultz2022} and \citet{owocki2022}, the incoherent radio luminosity $L_\mathrm{rad}$ correlates with the centrifugal breakout luminosity $L_\mathrm{CBO}$ obtained under the assumption that the magnetic field behaves as a monopole at the reconnection site. In that case, the CBO luminosity is given by\footnote{Note that the expression reported in \citet{owocki2022} has $B_\mathrm{d}$ in place of $B_\mathrm{eq}$, which was a typographical error.}:
\begin{align}
    L_\mathrm{CBO}&=\frac{B_\mathrm{eq}^2R_*^4\Omega^2}{v_\mathrm{orb}},\label{eq:Lcbo_expression}
\end{align}
where $\Omega=2\pi/P_\mathrm{rot}$ is the angular rotation speed, $v_\mathrm{orb}=\sqrt{GM_*/R_*}$ is the surface orbital speed and $G$ is the universal gravitational constant, $B_\mathrm{eq}=B_\mathrm{d}/2$ is the surface magnetic field strength at the magnetic equator. 
This relation is approximately equivalent to the empirical relation of \citet{leto2021} that differs by the ratio between stellar mass and radius ($M_*/R_*$), which is nearly constant for main-sequence stars.
In order to investigate the energy reservoir of the total incoherent, non-thermal radio power, we estimate the incoherent radio luminosities for the newly detected radio-bright magnetic hot stars. 

Note that while our expression for $L_\mathrm{CBO}$ is identical to that adopted by \citet{owocki2022} while deriving the empirical relation between $L_\mathrm{CBO}$ and $L_\mathrm{rad}$, \citet{keszthelyi2025} used a slightly different expression for $L_\mathrm{CBO}$ that involves $B_\mathrm{d}$ in place of $B_\mathrm{eq}$ (see their Equation 5). However, the $L_\mathrm{CBO}$ values obtained using these two different expression differ only by a constant factor of $4$.

In the following subsections, we present our strategy to estimate the incoherent radio luminosities and the correlation relation with the expanded sample.

\begin{figure}
    \centering
    \includegraphics[width=\columnwidth]{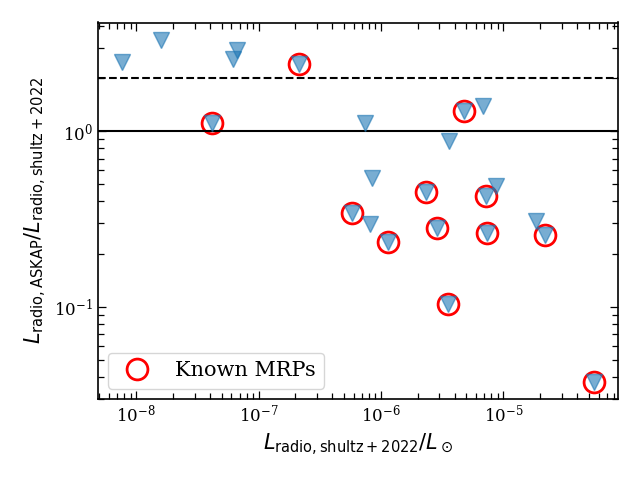}
    \caption{Comparison of radio luminosity estimated using only the ASKAP measurements to that reported by \citet{shultz2022}. See \S\ref{sec:results} for a description of the procedure to estimate radio luminosity from spectral radio luminosity. The known MRPs are highlighted with red unfilled circles. The horizontal solid and dashed lines mark $L_\mathrm{radio, ASKAP}=L_\mathrm{radio,shultz+2022}$ and $L_\mathrm{radio, ASKAP}=2\times L_\mathrm{radio,shultz+2022}$ respectively, where $L_\mathrm{radio, ASKAP}$ is the radio luminosity obtained using ASKAP measurements only, and $L_\mathrm{radio, shultz+2022}$ is the radio luminosity from \citet{shultz2022}.
    }
    \label{fig:Lrad_comparison}
\end{figure}

\subsubsection{Estimating incoherent radio luminosity}\label{subsec:Lrad_estimation}
To calculate the incoherent radio luminosity, wideband observations are required to perform an integration over frequencies. However, only ten stars have wideband observations suitable for spectral modelling. These stars exhibit a decline in their flux densities below $\sim 1$ GHz, and flat spectra between 1 and $\sim$ tens of GHz frequencies \citep{leto2021}. Based on these observations, \citet{shultz2022} adopted the strategy of integrating a trapezoidal function between 0.6--100 GHz, with zeros at the extrema and a flat spectrum between 1.5--30 GHz, with the peak flux density set to the highest incoherent flux density observed at any frequency from that star. 

For the newly added radio-bright magnetic hot stars, we employ this strategy with a slight modification, we extend the flat part of the spectra down to our lowest observing frequency. This was 0.9 GHz for most cases, which changes the luminosities by a factor of 1.005 as compared to those estimated following the strategy of \citet{shultz2022}.
For the stars that were already included in \citet{shultz2022}, we use the reported radio luminosities except for a few cases (\S\ref{subsec:Lrad_comparison}). 

\begin{figure*}
    \centering
    \includegraphics[width=0.49\textwidth]{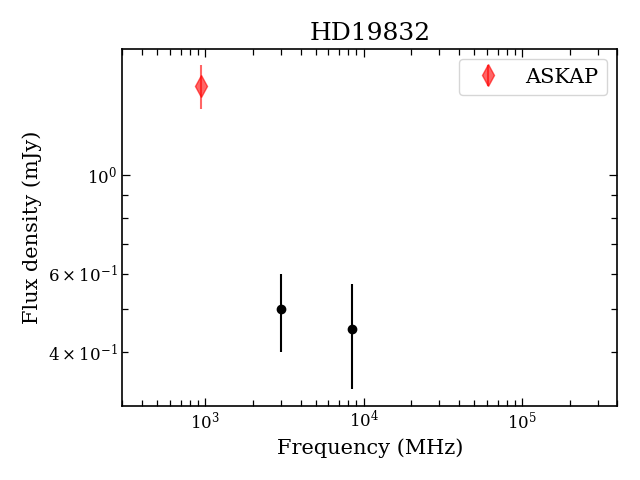}
    \includegraphics[width=0.49\textwidth]{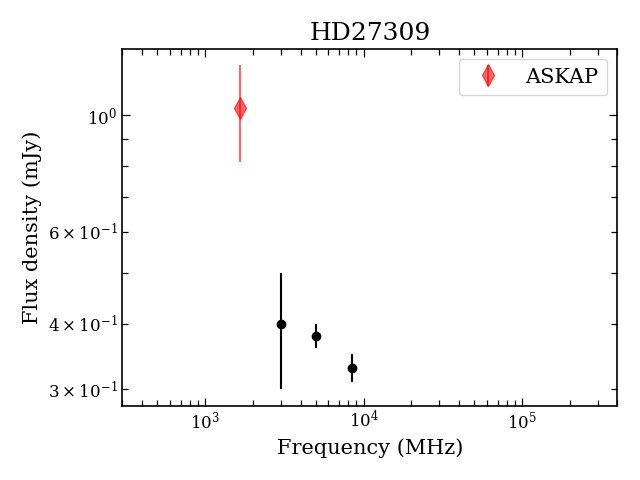}
    \includegraphics[width=0.49\textwidth]{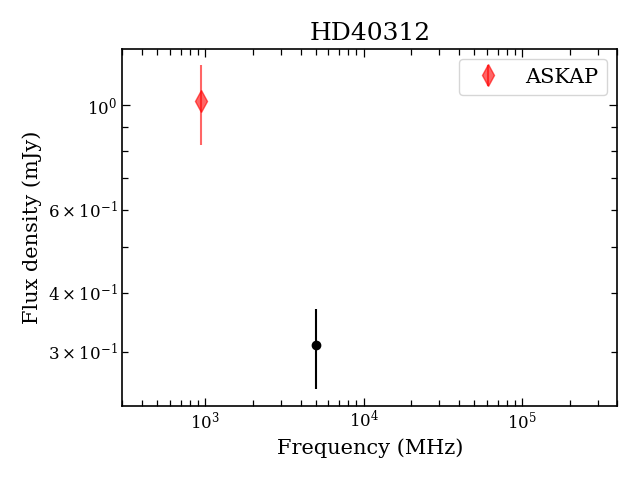}
    \includegraphics[width=0.49\textwidth]{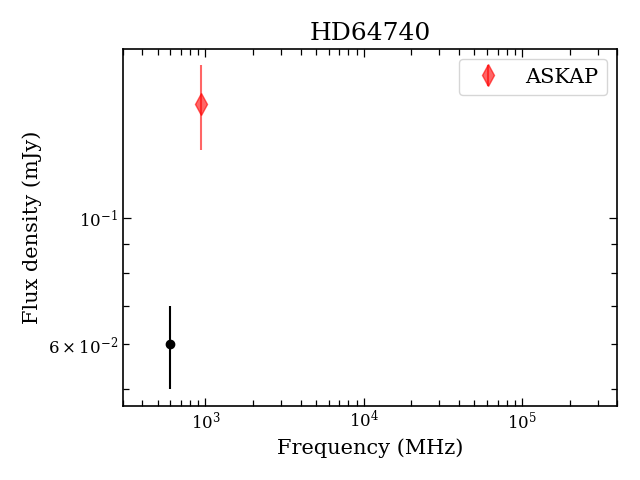}
    \includegraphics[width=0.49\textwidth]{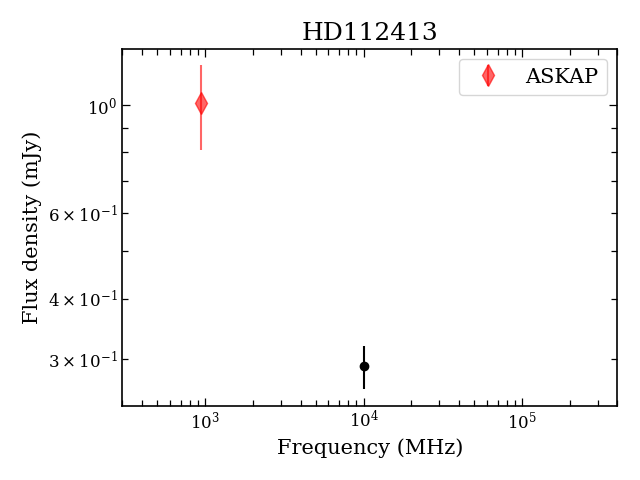}
    \caption{Spectra of ASKAP detected stars already included in the sample of \citet{leto2021} and \citet{shultz2022}, for which the radio luminosity estimated using only the ASKAP flux density measurements are higher by a factor $\geq 2$ than that reported by \citet{shultz2022}. 
    For all of these stars, the flux density measured by ASKAP is higher than the flux densities reported at other wavebands \citep[used by][]{shultz2022}. Besides, barring HD\,64740, the combined measurements violate the assumption on the spectral shape made by \citet{shultz2022} while estimating incoherent radio luminosity.}
    \label{fig:common_star_spectra_outlier}
\end{figure*}

While the fact that both coherent and incoherent radio emission are observable at ASKAP band \citep[e.g.][etc.]{das2022,pritchard2021} enhances the versatility of our project, it also implies that more caution is needed to avoid misidentifying the emission mechanism. This problem will be resolved as more epochs of data are accumulated as one will easily be able to identify the coherent radio emission based on variability alone. At the current stage, we adopt the following strategy for the newly detected radio stars in order to minimise the chance of wrongly using coherent radio flux density in the estimation of incoherent radio luminosity:
\begin{enumerate}
    \item For stars with multi-epoch detections, we use the minimum flux density observed. This is justified because the rotational modulation of incoherent radio emission is much less prominent at $\lesssim 1$ GHz \citep[e.g.][]{leto2018,das2018,lim1996}. In addition, taking the minimum flux density further minimises the probability of contamination from coherent emission.
    \item We use only those flux densities (total intensity) for which the corresponding  circular polarisation percentage is $<20\%$. Note that no detectable circular polarisation in the incoherent radio emission has been reported from magnetic hot stars at 700 MHz, while it has been reported at a level of $\sim 20\%$ at 44 GHz \citep{leto2017}.
\end{enumerate}
Two of the 49 stars did not satisfy the above criteria at any epoch. Both these stars (HD\,142990 and HD\,196178) are already included in the sample of \citet{shultz2022} and we used the radio luminosity reported by \citet{shultz2022} for these two stars. 

\subsubsection{Comparison with estimates from \citet{shultz2022}}\label{subsec:Lrad_comparison}
After excluding HD\,142990 and HD\,196178, we have 23 stars common to our sample, i.e. both detected by ASKAP and included by \citet{shultz2022}.
For this sub-sample, we compare the incoherent radio luminosities obtained following the above strategy ($L_\mathrm{rad,ASKAP}$) with those reported ($L_\mathrm{rad,Shultz+2022}$).
Since the incoherent radio emission appears to be more prominent above 1 GHz \citep{leto2021}, we expect the radio luminosities estimated by using only the ASKAP data (at $\approx 1$ GHz) to represent a lower limit to the true radio luminosity. 

The comparison is shown in Figure \ref{fig:Lrad_comparison}. For the majority of the stars, $L_\mathrm{rad,ASKAP}\lesssim L_\mathrm{rad,Shultz+2022}$ as expected. However for five stars, $L_\mathrm{rad,ASKAP}\geq 2\times L_\mathrm{rad,Shultz+2022}$. These are HD\,19832, HD\,27309, HD\,40312, HD\,64740 and HD\,112413.
In order to investigate the underlying reason, we plot the radio spectrum (Figure \ref{fig:common_star_spectra_outlier}) for each of these stars by combining the flux density measured by ASKAP that was used in the estimation of incoherent radio luminosity, with those already considered by \citet{leto2021} and \citet{shultz2022}. 
We find that barring HD\,64740, the flux densities of the remaining stars decrease with increasing frequencies between $1-10$ GHz, which violates the assumption made when estimating the incoherent radio luminosity. In the case of HD\,64740, \citet{shultz2022} estimated the incoherent radio luminosity based on its flux density measurement at $600$ MHz (the only existing measurement at the time) and found the star to be underluminous in radio. 
Similar to HD\,64740, the previous estimation of incoherent radio luminosities for HD\,40312 and HD\,112413 were based on measurement at single frequencies \citep[though at frequencies higher than the ASKAP observations,][]{shultz2022}.
Considering that for these three stars, we do not have a reason to favour the estimations of \citet{shultz2022} over our own, we choose to use the ASKAP measurements to estimate the incoherent radio luminosities.


In the cases of HD\,27309 and HD\,19832, the ASKAP flux densities are significantly higher than those at higher frequencies. HD\,19832 is a known MRP \citep[discovered at 700 MHz]{das2022}. Thus, for this star, it is very likely that the flux density observed at $\approx 1$ GHz has a significant contribution from the coherent component despite the low circular polarisation. The same possibility cannot be ruled out for HD\,27309 even though it is not a known MRP yet. As a result, for these two stars, we use the incoherent radio luminosities reported by \citet{shultz2022} instead of using the ASKAP data.


The spectra of the remaining stars are provided in Appendix (Figure \ref{fig:common_star_spectra_normal}). We use the incoherent radio luminosities reported by \citet{shultz2022} for these stars except for the cases of HD\,35298, HD\,61556 and HD\,105382 that exhibit `peculiar' properties. The reason(s) for favouring our own estimations over those reported by \citet{shultz2022} are described in \ref{sec:peculiar_spectra}.

Figure \ref{fig:Lrad_comparison} also suggests that the ratio between the two luminosities $L_\mathrm{rad,ASKAP}/ L_\mathrm{rad,Shultz+2022}$ decreases with increasing $L_\mathrm{rad,Shultz+2022}$. Since the ASKAP luminosities involve flux densities at $\approx 1$ GHz and the luminosities reported by \citet{shultz2022} primarily involve flux densities above 1 GHz, the ratio is a measure of radio spectral index over $\lesssim 1$ GHz to $>1$ GHz frequencies. Thus, the radio spectral index seems to have a dependence on radio luminosities. 
To investigate this aspect systematically, it will be important to adequately sample the radio spectra in order to obtain meaningful estimates of spectral indices and precise estimations of incoherent radio luminosities.

\begin{figure*}
\centering
    \includegraphics[width=0.97\textwidth]{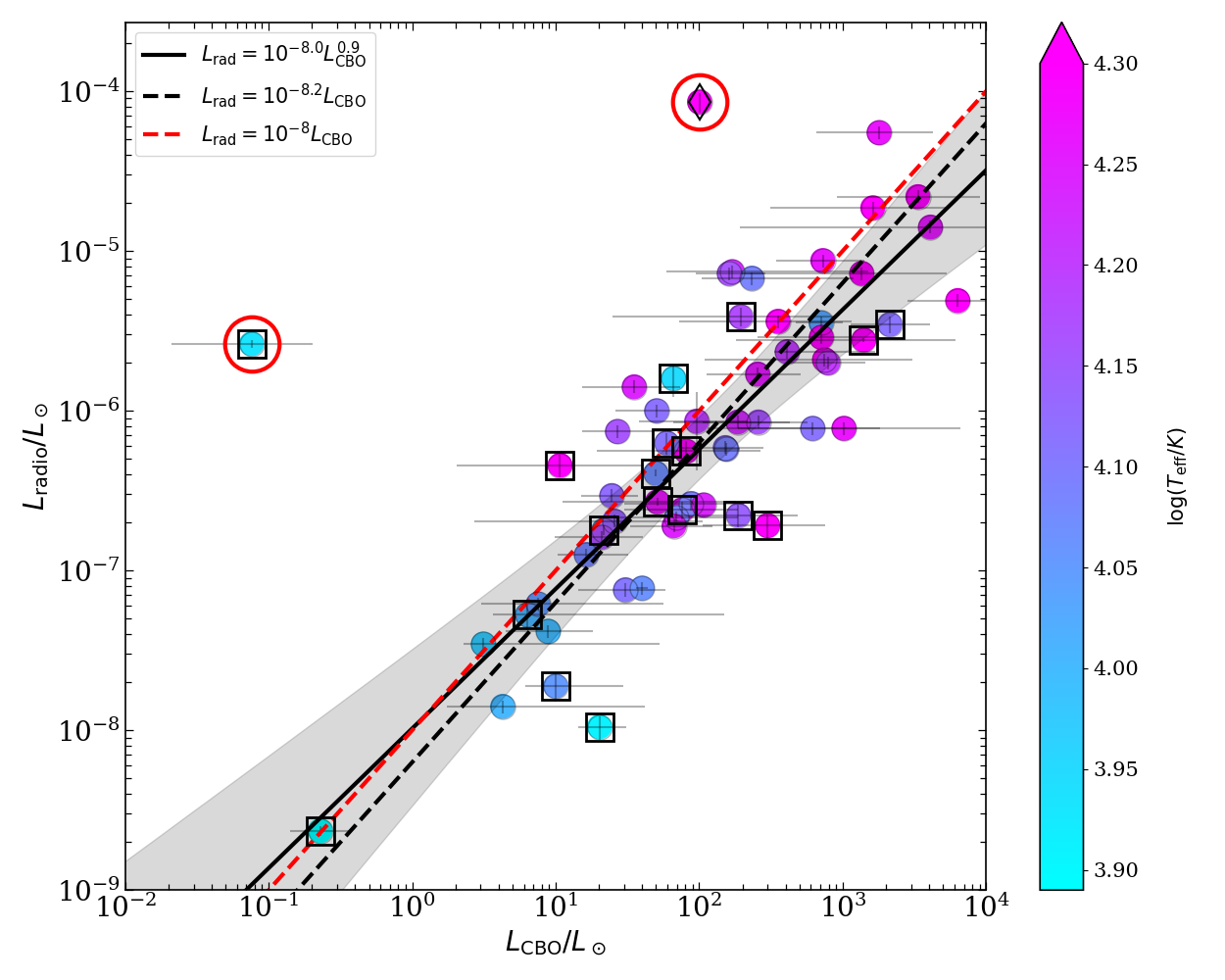}
    \caption{The correlation between non-thermal radio luminosity and CBO luminosity. The stars marked with squares represent the stars for which the incoherent radio luminosity is obtained using ASKAP flux density measurements. The star enclosed in diamond is HD\,148937, the only star in a sample that lacks a CM. This star and HD\,101412 are the two outliers (overluminous) and are marked with red circles.
    The solid black line represents the best-fit relation between the two quantities. The shaded region marks the $3\sigma$ deviation from the best-fit values. The black dashed line shows the best-fit line if we use a function of the form $L_\mathrm{rad}=10^{-\alpha}L_\mathrm{CBO}$.
    Red dashed line shows the empirical relation reported by \citet{shultz2022}.
    In colour scale, we also show the effective temperature of the stars. 
    \label{fig:Lrad_LCBO}
    }
\end{figure*}
\subsubsection{The correlation relation between incoherent radio luminosity with CBO luminosity}\label{subsec:correlation}
\citet{owocki2022} reported a correlation between incoherent radio luminosity $L_\mathrm{rad}$ and the CBO luminosity $L_\mathrm{CBO}$ (given by Equation \ref{eq:Lcbo_expression}) of the form $L_\mathrm{rad}=10^{-8}L_\mathrm{CBO}$. Most recently,\citet{keszthelyi2024} revisited the correlation (with the addition of a new star discovered at 650 MHz) and reported a slightly modified form: $L_\mathrm{rad}=10^{-8.52}L_\mathrm{CBO}^{0.88}$.
After taking into account of the difference in the mathematical expression used for $L_\mathrm{CBO}$ (Section \ref{subsec:result_incoherent}), this relation translates to $L_\mathrm{rad}=10^{-8.0}L_\mathrm{CBO}^{0.88}$.

For the newly added stars, only a subset has all the necessary information required to calculate $L_\mathrm{CBO}$. 
In Figure \ref{fig:Lrad_LCBO}, we plot the $L_\mathrm{rad}$ against $L_\mathrm{CBO}$ for the expanded sample. The newly added stars extend the range of $L_\mathrm{CBO}$ by an order of magnitude. 
There are two clear outliers shown with red circles. Neither of these stars was included in the sample of \citet{shultz2022} and \citet{leto2021}. One of the stars is HD\,148937, the only radio-bright star without a CM. Thus, the CBO scenario is not relevant for this star. 
The other star is HD\,101412 that harbours a CM. 
Possible reasons for its deviation will be discussed in \S\ref{subsec:scatter}.
In the subsequent discussion, we exclude both HD\,148937 and HD\,101412 from our sample.

The key challenge in quantifying the relation between $L_\mathrm{rad}$ and $L_\mathrm{CBO}$ is the lack of reliable uncertainties associated with the radio luminosities. The error bars on $L_\mathrm{rad}$ (in Figure \ref{fig:Lrad_LCBO}) likely underestimate the true uncertainties due to lack of spectral information and also not considering the variability of flux density with rotational phase.
Nevertheless, we fit a straight line between $\log_{10}L_\mathrm{CBO}$ and $\log_{10}L_\mathrm{rad}$, without including the error bars, by performing a Markov Chain Monte Carlo (MCMC) analysis using the \texttt{lmfit} package \citep{newville2014}.
We set the parameter \texttt{is\_weighted} to \texttt{False}\footnote{\url{https://emcee.readthedocs.io/en/stable/tutorials/line/}}.
In this case, the function also returns an estimate of the true uncertainty in the data ($L_\mathrm{rad}$). We vary the slope between $0.5$ and $1.2$ and the intercept between $-10$ and $-7$. The best-fit value of the slope $m$ comes out to be $0.87$ with a $1\sigma$ confidence interval of $[0.80,0.94]$. The best-fit value of the intercept is $-8.0$ with a $1\sigma$ confidence interval of $[-8.1,-7.8]$. These values are consistent with those obtained by \citet{keszthelyi2025} after accounting for the difference in the definition of $L_\mathrm{CBO}$. The intercept also matches that reported by \citet{owocki2022}, though the slope is slightly smaller than unity.
The true uncertainty in $\log_{10}L_\mathrm{rad}$ is estimated to be $0.47$ with a $1\sigma$ confidence interval of $[0.43,0.52]$.

If we fix the slope at unity, the best-fit intercept comes out to be $-8.2$ with a $1\sigma$ confidence interval of $[-8.31,-8.18]$. Thus, the expanded sample suggests slightly lower efficiency in radio production than that reported by \citet{owocki2022}. The different empirical relations are shown in Figure \ref{fig:Lrad_LCBO}.

\subsubsection{Investigating the role of temperature}\label{subsec:partial_corr}
From Figure \ref{fig:Lrad_LCBO}, it appears that the majority of the hotter stars have higher $L_\mathrm{rad}$ and also higher $L_\mathrm{CBO}$. 
In order to disentangle the effects of temperature and CBO luminosities, 
we examine the partial correlation coefficients.
The partial correlation coefficient is an indicator of the degree of correlation between two quantities $x$ and $y$, after removing their dependence on a third quantity $z$. In other words, it allows us to test whether an observed correlation between $x$ and $y$ is a result of their individual dependence on a third quantity $z$. 
To calculate the partial correlation coefficients, we use the \texttt{Pingouin} package by \citet{vallat2018}.
Taking $L_\mathrm{CBO}$ as $x$, $L_\mathrm{rad}$ as $y$ and $T_\mathrm{eff}$ as $z$, the partial correlation coefficient comes out to be 0.75 with a $95\%$ confidence interval of (0.61,0.85) and a p-value of $10^{-11}$. A small p-value indicates a statistically significant correlation and vice-versa. 
This shows that $L_\mathrm{CBO}$ and $L_\mathrm{rad}$ are strongly correlated even after removing any possible dependence on temperature. If we choose to use semi-partial correlation coefficient, i.e. remove the effect of $z$ ($T_\mathrm{eff}$) on only one of the variables, the correlation coefficients come out to be somewhat lower, but still higher than 0.5 with small p-values ($\lesssim 10^{-6}$).

If we swap $x$ and $z$, i.e., examine whether there is any correlation between $L_\mathrm{rad}$ and $T_\mathrm{eff}$ after removing the effect of $L_\mathrm{CBO}$, the resulting correlation coefficient comes out to be $\sim 0.1$.
Thus, with the current data, there is no evidence of a role of $T_\mathrm{eff}$ in driving incoherent radio emission.

We also examine the correlation between $L_\mathrm{CBO}$ and $T_\mathrm{eff}$. The Spearman rank correlation coefficient between the two quantities comes out to be $0.6$ with a p-value of $\sim10^{-7}$, suggesting a statistically significant positive correlation. However, the correlation strength reduces to 0.3 (with a p-value of $0.03$) when we use partial correlation coefficient with stellar mass as the third ($z$) variable. In other words, the moderate correlation between $L_\mathrm{CBO}$ and $T_\mathrm{eff}$ arises due to the intrinsic dependencies of stellar radius (contained in $L_\mathrm{CBO}$) and effective temperature on stellar mass for main-sequence stars \citep[e.g.][]{eker2018}.

To summarise, we have strong evidence in favour of $L_\mathrm{CBO}$ being the true quantity governing the incoherent radio luminosity.

\subsection{Coherent radio emission}\label{subsec:ecme}
As mentioned already, we identify MRP candidates using two properties of ECME: (a) circular polarisation and (b) variability. Using the cross-matching method in Section\,\ref{sec: cross-match results}, we find nine magnetic hot stars with $|V|/I>15\%$ (Table\,\ref{tab: stokes V star results}), eight of which satisfy the criteria $|V|/I>30\%$. Among them, two are known MRPs (HD\,61556 and HD\,142301). The remaining six are new MRP candidates.

\begin{figure*}
    \includegraphics[width=0.45\textwidth]{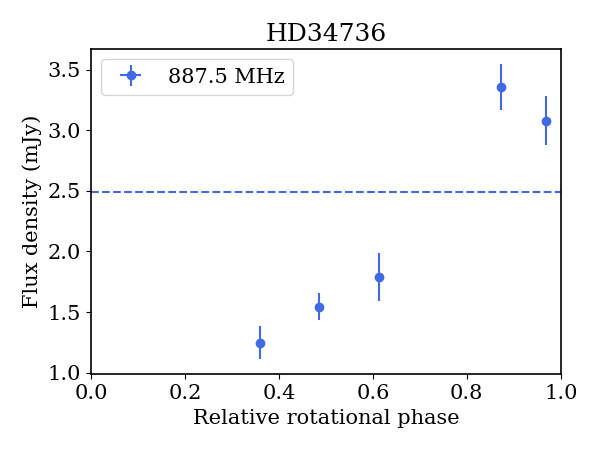}
    \includegraphics[width=0.45\textwidth]{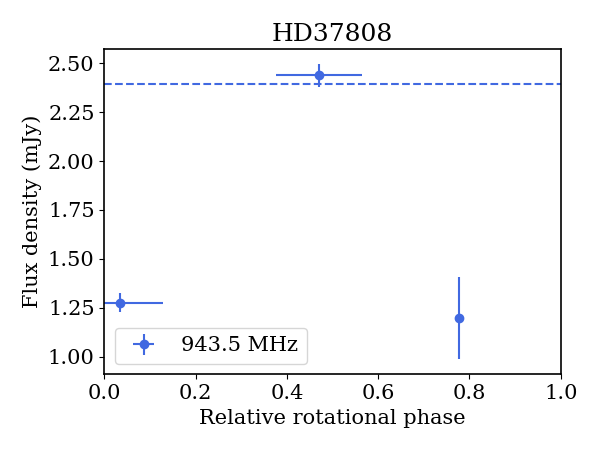}
    \includegraphics[width=0.45\textwidth]{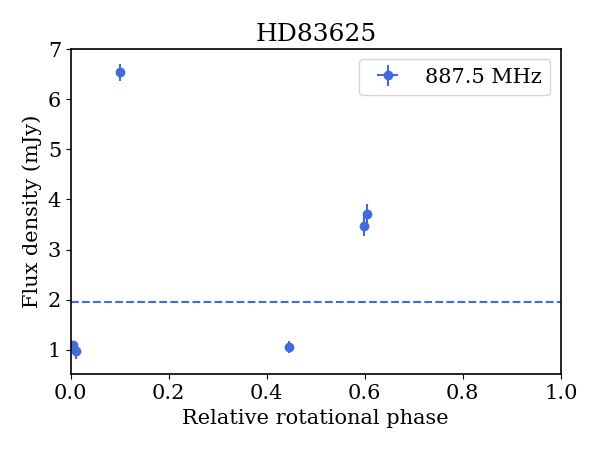}
    \includegraphics[width=0.45\textwidth]{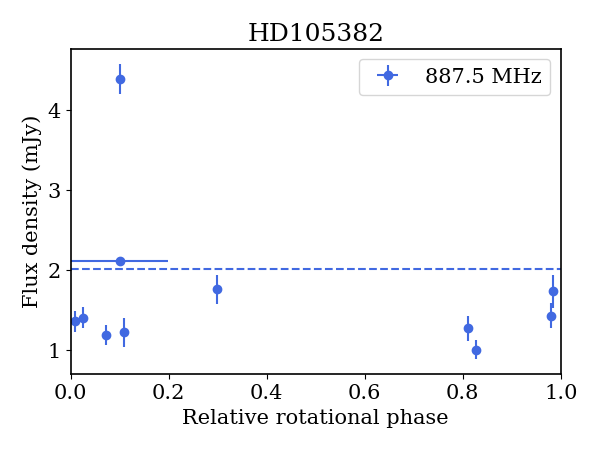}
    \includegraphics[width=0.45\textwidth]{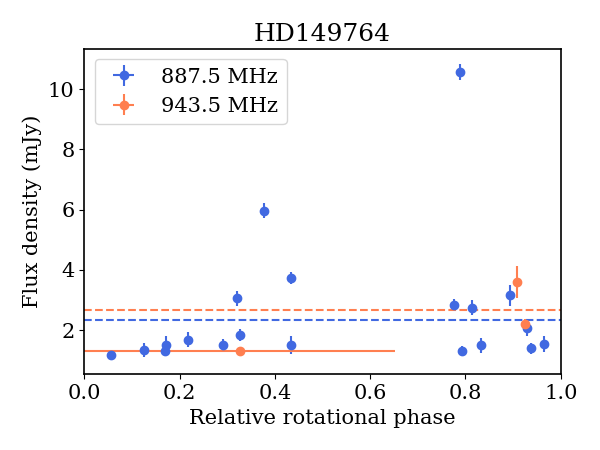}
    \includegraphics[width=0.45\textwidth]{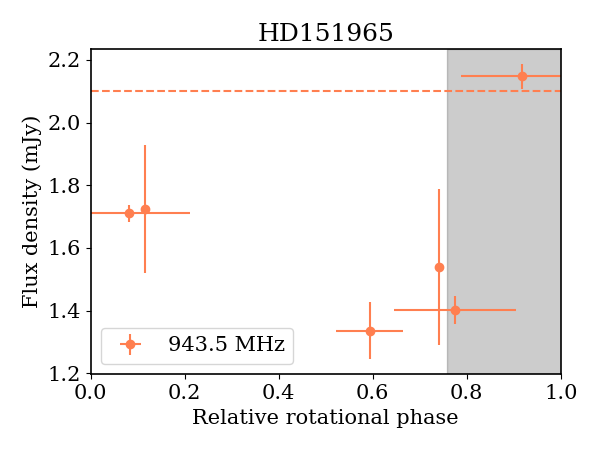}
    \caption{Light curves of the MRP candidates identified based on the variability in their flux densities. Note that only the light curves that satisfy at least one of the criteria listed in \S\ref{subsubsec:coherent} are shown here. Except for HD\,151965, the light curves for the rest satisfy the criterion that the flux density changes by a factor $>2$ between the different epochs of observations; the horizontal dashed lines mark the flux density twice the minimum observed flux density. In the case of HD\,151965, the light curve at 943.5 MHz satisfies the third criterion listed in \S\ref{subsubsec:coherent}. Here, the flux density changes by a factor greater than 1.5 (indicated by the dashed horizontal line) within a rotational phase window of width 0.16 about the phase of maximum flux density. The grey shaded region marks $\pm0.16$ phase around the phase of maximum flux density.}\label{fig:mrp_candidate_lightcurves}
\end{figure*}

Applying the two criteria related to variability (\S\ref{subsubsec:coherent}) resulted into $10$ MRP candidates. These include HD\,124224, HD\,12447, HD\,133880 and HD\,61556, all of which are known MRPs. The remaining six include HD\,34736, HD\,37808 and HD\,105382, which also satisfy the high circular polarisation criterion. The new MRP candidates obtained solely from the variability criteria are HD\,83625, HD\,149764 and HD\,151965. The light curves for all the six MRP candidates are shown in Figure \ref{fig:mrp_candidate_lightcurves}.

All the nine MRP candidates and the criteria based on which they are identified are listed in Table \ref{tab:mrp_candidates}. 
Three of these MRP candidates (HD\,83625, HD\,105382 and HD\,149764) have been confirmed to be MRPs using follow-up observations with the Australia Telescope Compact Array \citep{das2025}. To examine whether the remaining candidates are MRPs or flaring stars, similar follow-up observations will need to be performed.

\begin{table}[]
    \centering
    \caption{MRP candidates reported in this paper. The first column provides the name of the star, the second column shows the criteria based on which they are identified; C.P. stands for circular polarisation and Var. stands for variability (see \S\ref{subsec:ecme}). The third and fourth columns provide the circular polarisation fraction and the maximum flux density enhancement observed respectively.}
\begin{tabular}{lllc}
\hline
Star & Criteria & ${\left|V\right|}/{I}$ & Enhancement factor \\
\hline
HD\,34736 & C.P., Var. & $0.66$ & $2.7$\\
HD\,37808 & C.P., Var. & $0.37$ & $2.0$ \\
HD\,83625 & Var. & NA & $6.7$ \\
HD\,105382 & C.P., Var. & $0.60$ & $4.4$ \\
HD\,122451 & C.P.& $0.40$ & NA \\
HD\,149764 & Var. & NA & $9.1$\\
HD\,151965 & Var. & NA & $1.6$\\
HD\,164224 & C.P. & $0.77$ & NA \\
HD\,196178 & C.P. & $0.50$ & NA \\
\hline
\end{tabular}
    \label{tab:mrp_candidates}
\end{table}

To summarise, our main results are:
\begin{enumerate}
    \item  Radio detections of 49 magnetic hot stars out of which \textbf{14} stars are new additions to the sample of radio-bright magnetic hot stars. With that, the sample size of such stars has increased to 70.
    \item Nine MRP candidates based on circular polarisation and variability criteria.
    \item The best-fit relation between the theoretical CBO luminosity and the incoherent radio luminosity is $L_\mathrm{rad}=10^{-8.52}L_\mathrm{rad}^{0.88}$, consistent with the result of \citet{keszthelyi2024} but in conflict with that of \citet{owocki2022}.
    \item Our sample has two outliers (neither included while obtaining the correlation relation). Both stars are significantly overluminous in terms of their incoherent radio emission than predicted by the CBO theory. One of them is HD\,148937, the only radio-bright magnetic hot star in the sample without a CM. While this aspect explains its deviation from the predicted model for stars with CMs, the same is not the case for the other star HD\,101412. Future multi-frequency observation will be able to provide insights in this regard.
   \item Although there appears to be a dependence on the effective temperature based on visual inspection of Figure \ref{fig:Lrad_LCBO},  using partial correlation coefficients, we show that the strong correlation between $L_\mathrm{CBO}$ and $L_\mathrm{rad}$ remains even after removing any possible correlation with temperature.
\end{enumerate}
\section{Discussion}\label{sec:discussion}
While our results prove the robustness of the newly proposed CBO scenario and its prediction for radio emission, there are several areas that are yet to be investigated. These are discussed in the following subsections.

\begin{figure}
    \centering
    \includegraphics[width=1.0\columnwidth]{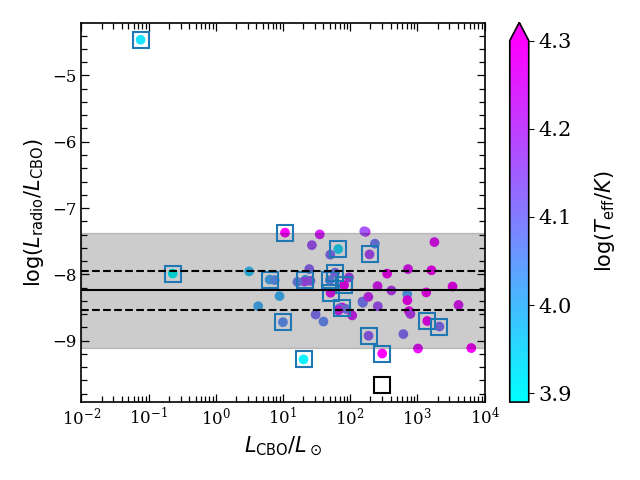}
    \caption{The ratio between log of observed incoherent radio luminosity to that predicted by the CBO theory. The solid horizontal line marks the median value, the dashed horizon lines represent the median absolute deviation (MAD) about the median value. The shaded regions corresponds to $3\times\mathrm{MAD}$. The empty square towards the bottom shows the position of HD\,64740, the hottest star in our sample, based on the incoherent radio luminosity reported by \citet{shultz2022}.\label{fig:Lrad_to_LCBO_ratio}}
\end{figure}

\subsection{Scatter in the correlation relation between incoherent radio luminosity and CBO luminosity}\label{subsec:scatter}
In Figure \ref{fig:Lrad_to_LCBO_ratio}, the difference between observed radio luminosity and that predicted by the CBO theory (monopole assumption) is examined for the expanded sample. The median value of the ratio is $L_\mathrm{rad}/L_\mathrm{CBO}\approx 10^{-8}$. We define a star to be underluminosus or overluminous if the corresponding value of $L_\mathrm{rad}/L_\mathrm{CBO}$ deviates from the median value by more than three times the median absolute deviations. Two stars are found to be underluminous (deviate by $\gtrsim 3.5\times\mathrm{MAD}$). One of these stars is HD\,64740, the hottest star in our sample (temperature of 25 kK), already speculated to be underluminous by \citet{shultz2022}. The other is HD\,148898, among the coolest stars in our sample (A-type, temperature of 8.1 kK). Considering that for both stars, the incoherent radio luminosity is determined by only using ASKAP data, it is not a surprise to find the resulting luminosity to be somewhat lesser than those expected (see \S\ref{subsec:Lrad_estimation}). 
It is worth noting that the deviation of HD\,64740 from the median relation was much higher when \citet{shultz2022} estimated the incoherent radio luminosity using the only flux density measurement available then, which was at 600 MHz (shown by the empty square in Figure \ref{fig:Lrad_to_LCBO_ratio}). 
In the future, observation at higher frequencies (and also lower for HD\,148898) will be useful to estimate their true incoherent radio luminosities and thus determine whether they are indeed underluminous or not. 
A quantitative analysis of how the deviation of the shape of the true spectrum from that assumed affects the estimated radio luminosities is provided in \ref{subsec:Lrad_test}.

The only star that deviates from the median by more than $4\times$MAD is HD\,101412. The star's estimated incoherent radio luminosity is $\approx 10^{-4.5}$ times its CBO luminosity.
This is a relatively cool star with unusually high radio luminosity and one of the newly added magnetic hot stars. The star was detected in the VAST survey at multiple epochs and was found to exhibit a nearly constant level of flux density. The lack of variability in the light curve and the lack of detection of circular polarisation (within measurement uncertainties) suggest that the detected emission represents incoherent radio emission (and not ECME). Thus, following \S\ref{subsec:Lrad_estimation}, its estimated radio luminosity should have been lower than predicted by the correlation relation.
The only unusual aspect about the star is its apparent spectral index inferred from its average flux density at 887.5 MHz and 943.5 MHz. The star was detected at 887.5 MHz at six epochs with the flux densities varying between 1.1 and 1.4 mJy. The average flux density comes out to be $1.2\pm0.1$ mJy. At 943.5 MHz, the star has only one flux density measurement: $0.77\pm0.04$ mJy, which indicates a spectral index of $-8.0\pm0.8$. Both the sign and magnitude of the spectral index are highly unusual for incoherent radio emission at these frequencies \citep[][also see Figure \ref{fig:common_star_spectra_normal} and \ref{fig:common_star_spectra_outlier}]{leto2021}. Note, however, that none of the 887.5 MHz measurements were acquired at the rotational phase corresponding to the measurement at 943.5 MHz.
Thus, the apparent high spectral index could actually be a result of variable flux density of the star. To confirm this aspect, this star should be observed over broadband (preferably simultaneously) and spanning a wider rotational phase range. This will allow us to determine the true shape of its incoherent radio spectrum, as well as to investigate temporal variability, and thus figure out whether it is truly `over-luminous' or not in terms of its incoherent radio emission.

In the sample of \citet{shultz2022}, HD\,171247 stood out as the only overluminous star. A number of reasons, such as binarity, stellar age, uncertainty in rotation period and contamination from coherent radio emission, were considered. In addition to these, it is worth noting that the star's incoherent radio luminosity was obtained using a single frequency measurement at 5 GHz. Thus, if HD\,171247 exhibits its incoherent radio emission predominantly at sub-GHz frequencies, its incoherent radio luminosity is actually less than that estimated under the assumption of a flat spectrum between 1.5 and 30 GHz.

Both VAST-MeMeS and the RAMBO projects \citep{keszthelyi2024} will be critical in constraining the lower frequency properties of incoherent radio emission.
Along with that, observation at mm-bands (above 43 GHz) will also be needed to constrain the upper turn-over frequency of the phenomenon and its dependence on stellar parameters.

In addition to obtaining more reliable estimates of radio luminosities, it is also important to ensure that the estimates of $L_\mathrm{CBO}$ are reliable (also see \ref{sec:LCBO_err}). \citet{keszthelyi2024} recently pointed out the need for long-term spectropolarimetric monitoring observations to fully characterise the stellar magnetic fields, as well as the consequence of large uncertainties in stellar rotation periods. In addition to magnetic field strength and rotation period, the other stellar parameter that strongly affects $L_\mathrm{CBO}$ is the stellar radius $R_*$. $L_\mathrm{CBO}$ has the strongest dependence on $R_*$: $L_\mathrm{CBO}\sim R_*^{4.5}$ (Equation \ref{eq:Lcbo_expression}).
Thus, even a $20\%$ uncertainty in $R_*$ leads to $\sim 100\%$ uncertainty in $L_\mathrm{CBO}$.

In the future, it will be important to overcome the above limitations so as to quantify the `true' scatter in the correlation relation. This will allow us to search for the role of other variables \citep[such as obliquity,][]{shultz2022}, and also to develop more insight on the apparent indifference to temperature or mass loss rates (\S\ref{subsec:indifference_to_Teff}).

\subsection{Possible reason behind indifference of radio emission on temperature}\label{subsec:indifference_to_Teff}
Intuitively, the incoherent radio luminosity is expected to have some dependence on stellar effective temperature (or mass-loss rate) due to two reasons. With increasing temperature, the strength of the stellar wind, which is the supplier of particles required for the production of radio emission, also increases. From this aspect, a positive correlation can be expected between incoherent radio luminosity and temperature. 
The subtle point, however, is that an increasing temperature or mass-loss rate will bring the reconnection sites closer to the stellar surface, and as a result, a reduction in the rotational kinetic energy of the wind materials, which in turn, reduces the energy available for the breakout events. Under the mono-pole assumption, these two competing effects exactly balance each other leaving no dependence on temperature or mass-loss rate.

There is, however, another way in which wind could be important.
A stronger wind also implies stronger free-free absorption (FFA) that will effectively reduce the observable incoherent radio luminosity. FFA was proposed to be the reason behind the seemingly weak radio emission from HD\,64740, the hottest star of the sample by \citet{shultz2022}. However, the discrepancy is reduced significantly following the use of a higher frequency flux density (in this case 944 MHz) instead of using that at 600 MHz \citep{shultz2022}. In this context, it is worth noting that above a temperature of 18--20 kK, coherent radio emission appears to be suppressed with further increase in temperature \citep[the proto-typical star $\sigma\,$ Ori E, an early-B type star, is a non-MRP,][]{leto2012,das2022c}. The current sample of radio-bright magnetic hot stars has only three stars hotter than $\sigma$ Ori E. To investigate the potential `quenching' of incoherent radio emission above a certain temperature, it will be important to acquire more radio observations of early-B and O-type stars with CM so as to either detect them or obtain deep upper limits.

In the future, we plan to use the upper limits (along with detections) obtained from the survey data to achieve this goal.

\section{Summary}\label{sec:summary}
The field of radio emission from magnetic early-type stars has recently undergone a revolution with the discovery of a scaling law connecting incoherent radio luminosity and stellar magnetospheric parameters, which was interpreted within the CBO scenario \citep{leto2021,shultz2022,owocki2022}. The concept of CBO is not new \citep[e.g. see Figure 4 of ][]{havnes1984}. However, it was recognised as the mode of plasma transport in magnetic hot stars only in 2020, based on properties of H$\alpha$ emission \citep{shultz2020,owocki2020}. The introduction of CBO to explain non-thermal radio emission is significant as, apart from providing a fundamentally new insights about how non-thermal radio is produced in large-scale stellar magnetospheres, this also offers a magnetospheric model that can consistently explain magnetospheric emission at two widely separated wavebands. 
The importance of this model only increases considering the suggestion that the same scaling law may also apply for much cooler UCDs and even planets \citep{leto2021}.

In view of the above, it is of utmost importance to scrutinise the above scaling law. The very first step to test its robustness is by increasing the sample size. In this very first paper of the `VAST-MeMeS' project that aims to characterise non-thermal radio emission, we examine the scaling law after adding 26 magnetic hot stars to the previous sample of 47 stars \citep{shultz2022}. We find that the scaling relation remains valid providing further support in its favour. 
We also note the importance of acquiring wideband radio emission to be able to observationally calculate incoherent radio luminosity. This will help us to investigate the role of other, potentially relevant stellar parameters in driving radio emission, as well as correlations between spectral properties (e.g. spectral indices, cur-off and turn-over frequencies etc.) and stellar magnetospheric parameters and the radio luminosity itself. 
Finally, it will be important to consider non-detections along with detections to fully characterise the emission \citep[demonstrated by,][]{shultz2022}, and also to explore whether incoherent radio emission is quenched (similar to coherent radio emission) due to absorption above certain temperature.

\newpage 
\begin{acknowledgement}
We thank the referee for their comments and suggestions that have helped us to improve our manuscript.
This scientific work uses data obtained from Inyarrimanha Ilgari Bundara, the CSIRO Murchison Radio-astronomy Observatory. We acknowledge the Wajarri Yamaji People as the Traditional Owners and native title holders of the Observatory site. CSIRO's ASKAP radio telescope is part of the Australia Telescope National Facility (\url{https://ror.org/05qajvd42}). Operation of ASKAP is funded by the Australian Government with support from the National Collaborative Research Infrastructure Strategy. ASKAP uses the resources of the Pawsey Supercomputing Research Centre. Establishment of ASKAP, Inyarrimanha Ilgari Bundara, the CSIRO Murchison Radio-astronomy Observatory and the Pawsey Supercomputing Research Centre are initiatives of the Australian Government, with support from the Government of Western Australia and the Science and Industry Endowment Fund.
BD acknowledges the Whadjuk people of the Noongar nation as the Traditional Owners of the land on which they perform the work.
LND acknowledges the Gadigal people of the Eora Nation as the traditional custodians of the land on which they perform most of their work.

KR thanks the LSST-DA Data Science Fellowship Program, which is funded by LSST-DA, the Brinson Foundation, and the Moore Foundation; Their participation in the program has benefited this work.
\end{acknowledgement}

\paragraph{Funding Statement}
We do not have any external funding to report.

\paragraph{Competing Interests}
None.

\paragraph{Data Availability Statement}
All the ASKAP data are publicly available via the CSIRO ASKAP Science Data Archive (CASDA, \url{https://research.csiro.au/casda/})
Table 6 is available in CSV format as a supplementary file.



\bibliography{hms.bib}

\appendix
\renewcommand{\thefigure}{A\arabic{figure}}

\setcounter{figure}{0}

\section{Peculiar radio spectra of magnetic massive stars}\label{sec:peculiar_spectra}
Among the stars for which $L_\mathrm{rad, ASKAP}\lesssim L_\mathrm{rad, Shultz+2022}$, three stars exhibit peculiar properties (Figure \ref{fig:common_star_spectra_normal}), for which we did not use their radio luminosities reported by \citet{shultz2022}.
The reason(s) for favouring our own estimations over those reported by \citet{shultz2022} are described below:

\textbf{HD\,35298:} HD\,35298 is an MRP discovered by \citet{das2019b}. The flux densities corresponding to ASKAP detection at around 1 GHz, and at 88 GHz are much higher than that at intermediate frequencies. The observed flux density at $\sim 1 $GHz (ASKAP measurement) is comparable to the peak flux density of coherent radio emission at these frequencies \citep{das2022b}, and is very likely of coherent origin\footnote{ECME from magnetic hot stars can have very low ($\sim$zero) circular polarisation \citep{das2022}. In other words, high circular polarisation is an indicator of ECME, but lack of high circular polarisation does not rule out ECME.}. The reason behind the excess flux density at mm-bands is unclear. We revise the incoherent radio luminosity of this star using the measurements at intermediate frequencies.

\textbf{HD\,61556:} HD\,61556 is another MRP, detected as a candidate by \citet{pritchard2021} and confirmed by \citet{das2022b}. For this star, \citet{shultz2022} used the flux density reported by \citet{pritchard2021} that had a circular polarisation of $76\%$, and thus extremely unlikely to represent incoherent radio emission. We hence use the ASKAP measurement that satisfies the criteria in \S\ref{subsec:Lrad_estimation} to estimate the incoherent radio luminosity.

\textbf{HD\,105382:} This star is not a confirmed MRP. However, the previous estimate of the incoherent radio luminosity by \citet{shultz2022} was based on the flux density measurement reported by \citet{pritchard2021} where they detected a percentage circular polarisation of 60\%. Such high circular polarisation indicates coherent radio emission suggesting that this star is also an MRP. We hence use new ASKAP measurement to revise its incoherent radio luminosity.

\section{Incoherent radio luminosity for stars with existing ultra wideband radio observations}\label{subsec:Lrad_test}
\begin{figure}
    \centering
    \includegraphics[width=0.95\linewidth]{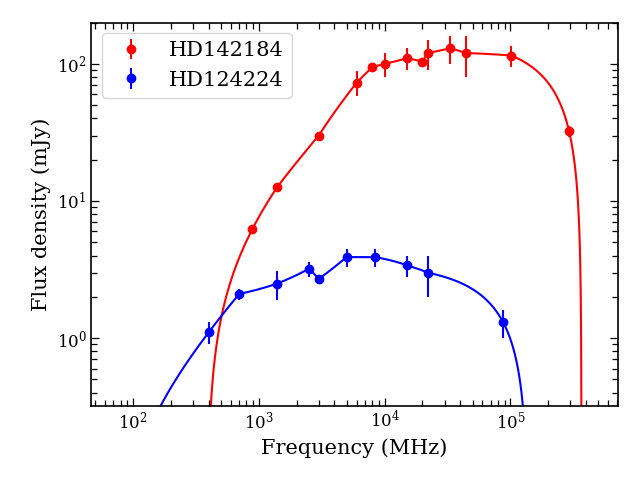}
    \caption{The incoherent gyrosynchrotron spectra of the only two hot magnetic stars for which spectral turn-over are observed on both ends of the spectra.}
    \label{fig:HD142184_HD124224}
\end{figure}

In our sample of stars, there are only two stars for which there are adequate data revealing the turn-over in the spectra of incoherent radio emission at both ends. These are HD\,124224 and HD\,142184. The spectra for these two stars are shown in Figure \ref{fig:HD142184_HD124224}. By using all the available measurements and integrating over a function indicated by the solid lines (Figure \ref{fig:HD142184_HD124224}), we find $\log (L_\mathrm{rad}/L_\odot)=-3.78\pm0.02$ and $-6.42\pm0.04$ for HD\,142184 and HD\,124224 respectively. The corresponding values estimated by \citet{shultz2022} are $-4.25\pm0.04$ and $-6.24\pm0.04$ respectively. Thus, the incoherent radio luminosity estimated for HD\,142184 by \citet{shultz2022} is much smaller than that obtained by using the actual spectrum, but that for HD\,124224 is very close to the one obtained by using the observed spectrum. This is not surprising since HD\,124224 approximately satisfies the assumptions about the spectral shape made by \citet{shultz2022}, whereas HD\,142184 strongly violates them (e.g. the spectrum does not turn over at 30 GHz and reach zero at 100 GHz). The discrepancy obtained for HD\,142184 is $0.47$ dex, which is very close to the uncertainty estimated by our MCMC analysis (\S\ref{subsec:correlation}).

\section{Estimating uncertainties in $L_\mathrm{CBO}$}\label{sec:LCBO_err}
\begin{figure}
    \centering
    \includegraphics[width=0.9\textwidth]{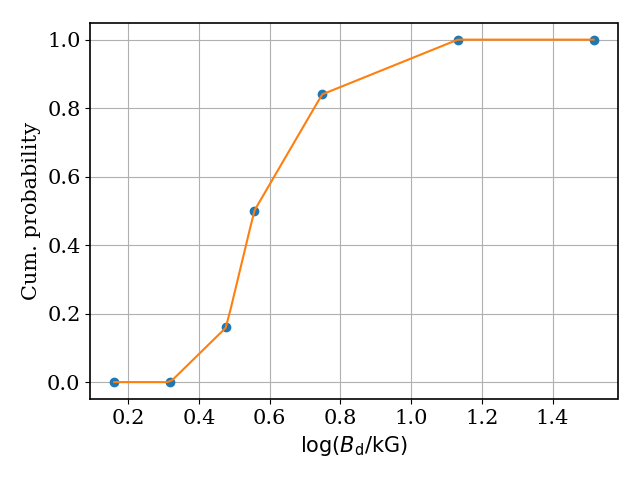}
    \includegraphics[width=0.9\textwidth]{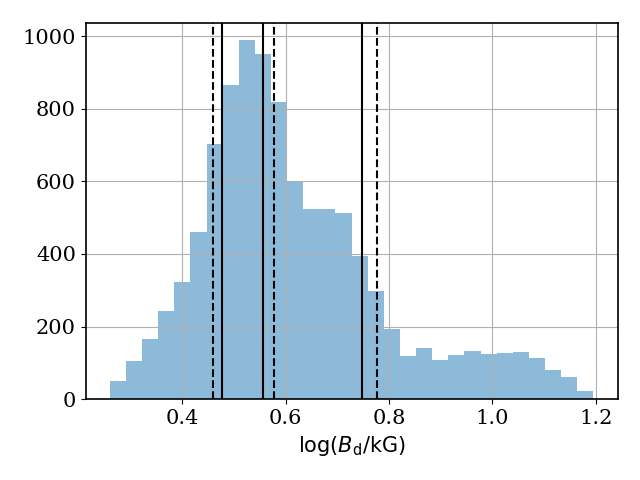}
    \caption{Demonstrating the construction of a probability distribution of $\log B_\mathrm{d}$ satisfying the available $B_\mathrm{d}$ measurements (\S\ref{sec:LCBO_err}). The \textbf{top} panel shows the cumulative probability distribution constructed following the strategy described in \S\ref{sec:LCBO_err}. The bottom panel shows the probability distribution of $\log B_\mathrm{d}$ obtained. The solid lines mark the `true' $16$, $50$ and $84$ percentiles (left to right respectively) and the dashed lines mark the same for the distribution.}
    \label{fig:prob_cont.}
\end{figure}

$L_\mathrm{CBO}$ is a function of stellar mass $M_*$, radius $R_*$, rotation period $P_\mathrm{rot}$ and polar magnetic field strength $B_\mathrm{d}$. Thus, the uncertainty in $L_\mathrm{CBO}$ is determined by the uncertainty in these quantities. However, for most of the stars, $B_\mathrm{d}$ has highly asymmetric error bars. This renders error propagation using analytical formula unfeasible. We hence adopted a Monte Carlo approach to obtain a representative error bars for the CBO luminosities. The main challenge here is to draw random values of $B_\mathrm{d}$ from a distribution that satisfies the available asymmetric uncertainties (assumed to be $1\sigma$). We achieve that approximately in the following way:

\begin{enumerate}
    \item Let the magnetic field strength value is $B_\mathrm{d}=B^{+b_u}_{-b_l}$.
    \item We construct the cumulative probability distribution for $\log B_\mathrm{d}$ by assigning probabilities of $0.16,\,0.50,\,0.84$ at $\log (B-b_l),\,\log B,\,\log(B+b_u)$ respectively.
    \item We assume that the cumulative probabilities are zero at $\log B-\left(n\times\{\log B-\log(B-b_l)\}\right)$, and unity for $\log B+\left(n\times\{\log(B+b_u)-\log B\}\right)$ for $n\geq 3$.
    \item The cumulative probabilities at other values of $\log B_\mathrm{d}$ are obtained by linear interpolation.
    \item The cumulative probability distribution constructed in this way is used to infer the probability distribution for the values of $\log B_\mathrm{d}$.
\end{enumerate}
An example of constructing the cumulative probability distribution and the resulting probability distribution in $\log B_\mathrm{d}$ is shown in Figure \ref{fig:prob_cont.}.

\begin{figure*}
\centering 
    \includegraphics[width=0.49\textwidth]{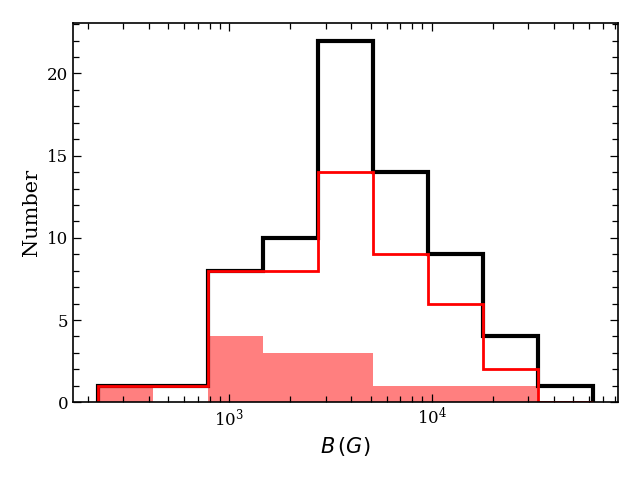}
    \includegraphics[width=0.49\textwidth]{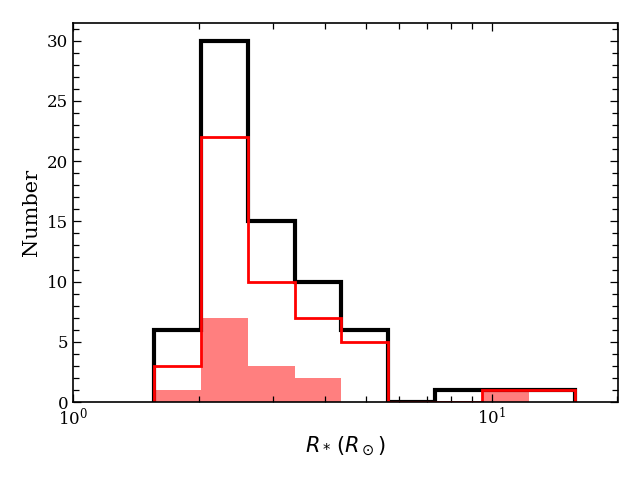}
    \includegraphics[width=0.49\textwidth]{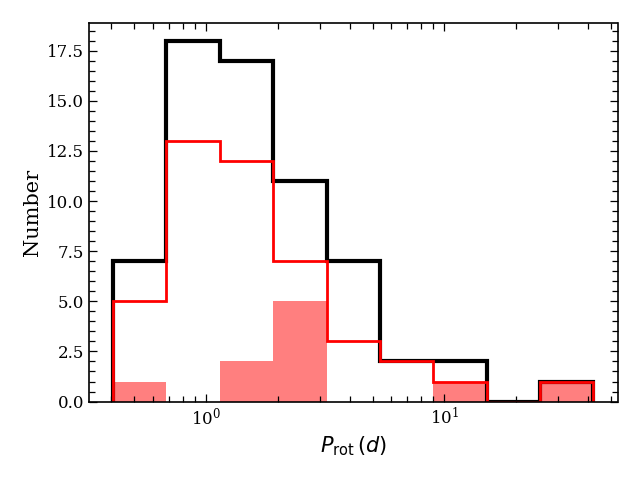}
    \includegraphics[width=0.49\textwidth]{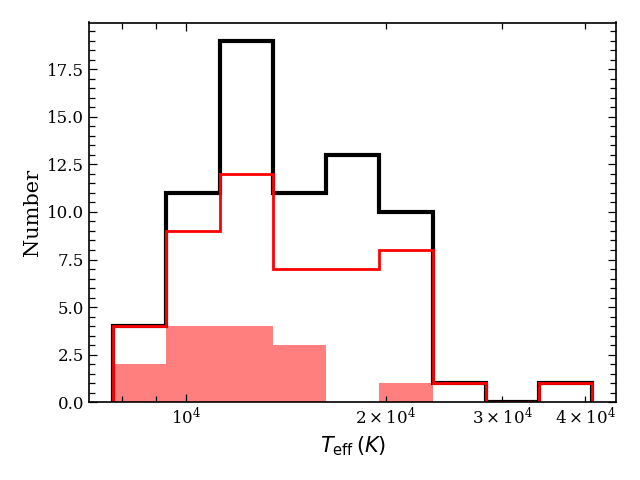}
\caption{The parameter span of the expanded radio-bright magnetic hot star sample. The black unfilled histograms correspond to the full sample of radio-bright magnetic hot stars (68 stars), the red unfilled histograms correspond to the magnetic hot stars detected by ASKAP (48 stars); and the red filled histograms correspond to the magnetic hot stars for which the ASKAP detections mark the first radio detection at any radio frequency (\textbf{14 stars}). Note that the number of magnetic hot stars for which the ASKAP detections mark the first radio detection within our observed frequency range is \textbf{24}.
\label{fig:parameter_span}}
\end{figure*}

\begin{figure*}
  \centering 
  \subfloat{\includegraphics[width=0.45\textwidth]{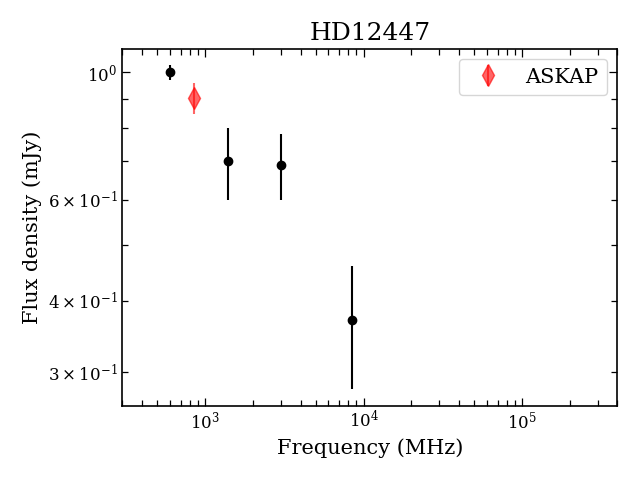}}
  \qquad 
  \subfloat{\includegraphics[width=0.45\textwidth]{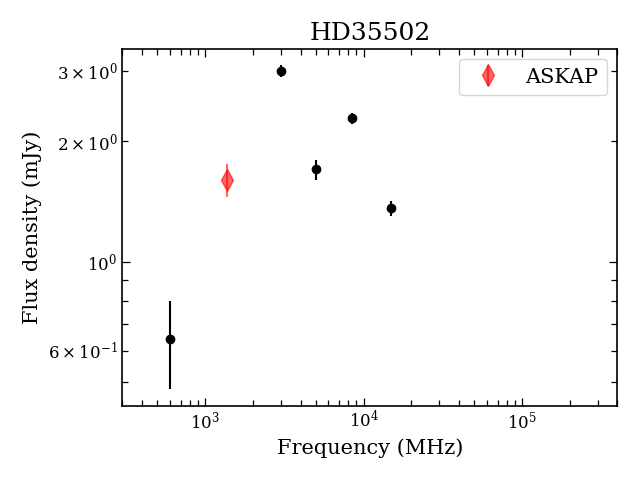}}
  \qquad 
  \subfloat{\includegraphics[width=0.45\textwidth]{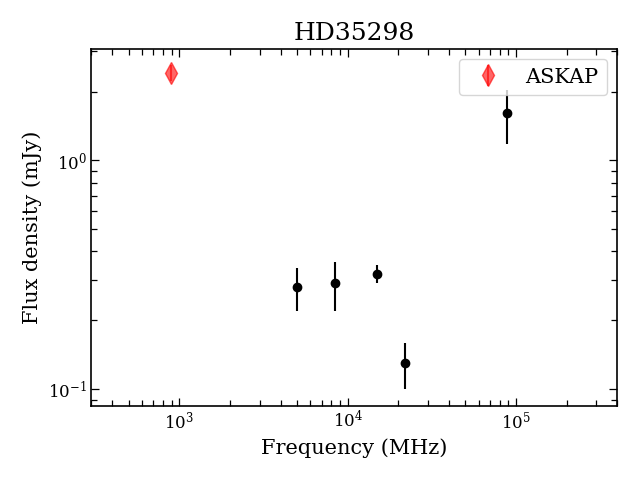}}
  \qquad 
  \subfloat{\includegraphics[width=0.45\textwidth]{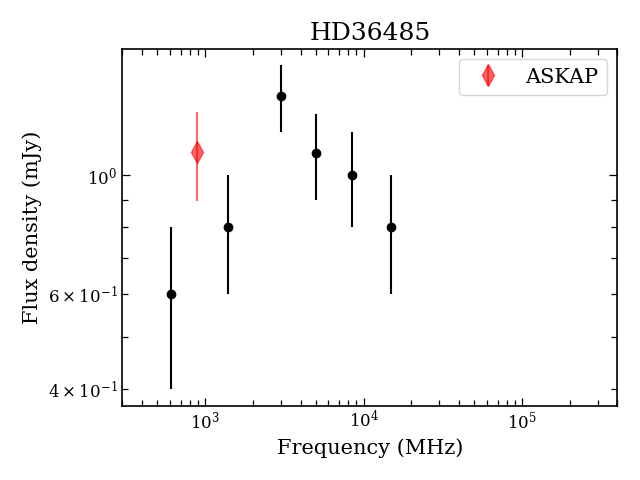}}
  \qquad 
  \subfloat{\includegraphics[width=0.45\textwidth]{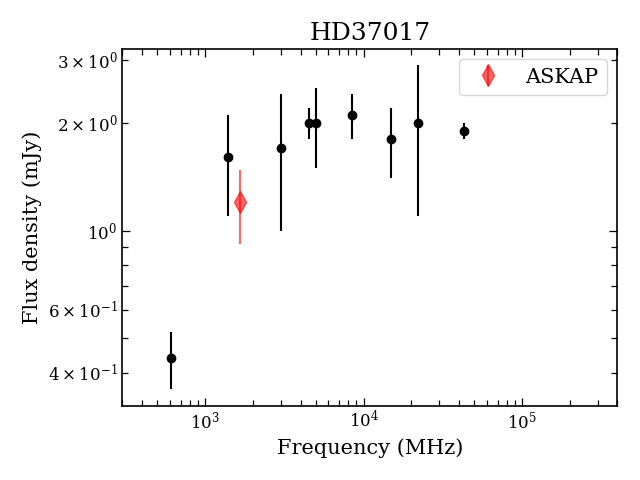}}
  \qquad 
  \subfloat{\includegraphics[width=0.45\textwidth]{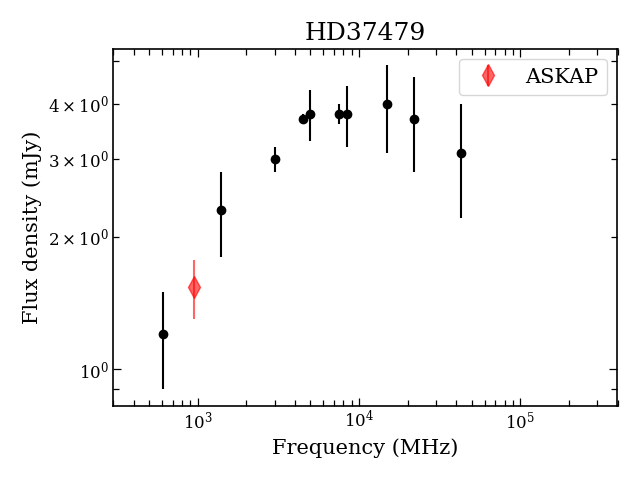}}
  \caption{Spectra of ASKAP detected stars already included in the sample of \citet{leto2021} and \citet{shultz2022}, for which the ratio of estimated radio luminosity estimated using only the ASKAP flux density measurements to that reported by \citet{shultz2022} are $\lesssim 2$.}
  \label{fig:common_star_spectra_normal}
\end{figure*}

\begin{figure*}
\ContinuedFloat 
  \centering 
  \subfloat{\includegraphics[width=0.45\textwidth]{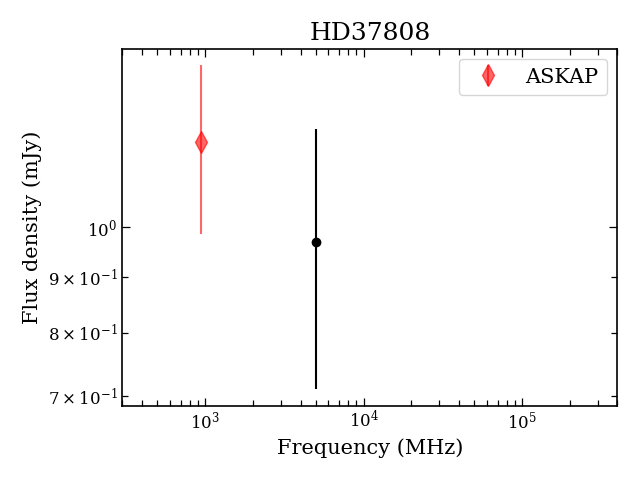}}
  \qquad 
  \subfloat{\includegraphics[width=0.45\textwidth]{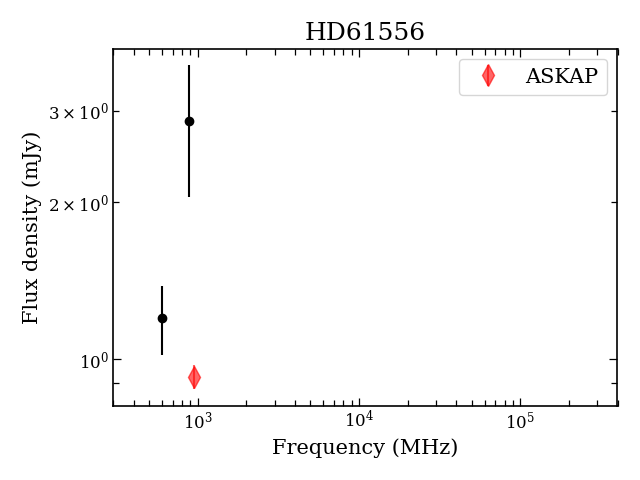}}
  \qquad 
  \subfloat{\includegraphics[width=0.45\textwidth]{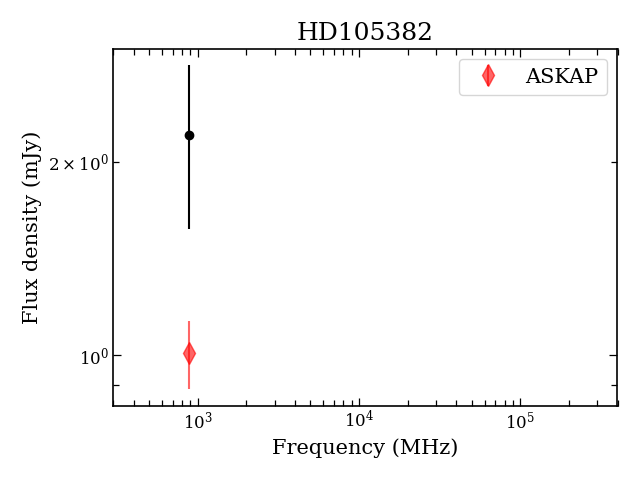}}
  \qquad 
  \subfloat{\includegraphics[width=0.45\textwidth]{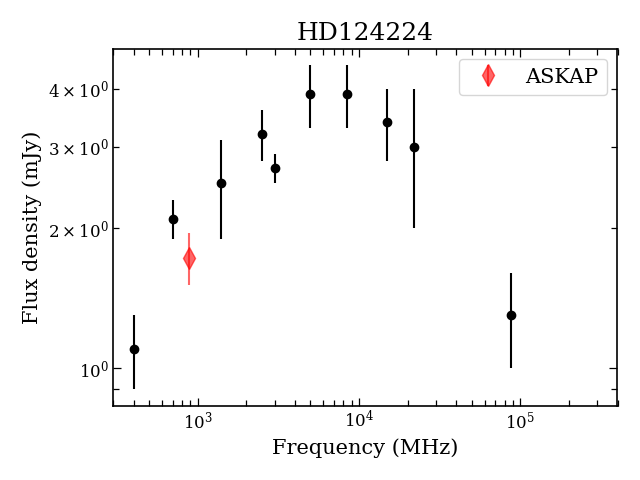}}
  \qquad 
  \subfloat{\includegraphics[width=0.45\textwidth]{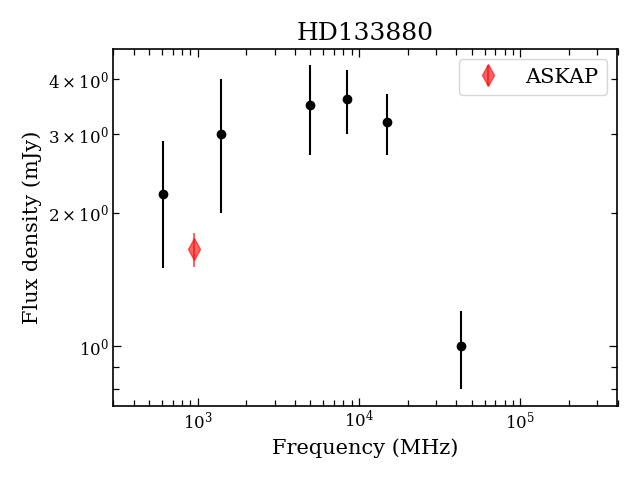}}
  \qquad 
  \subfloat{\includegraphics[width=0.45\textwidth]{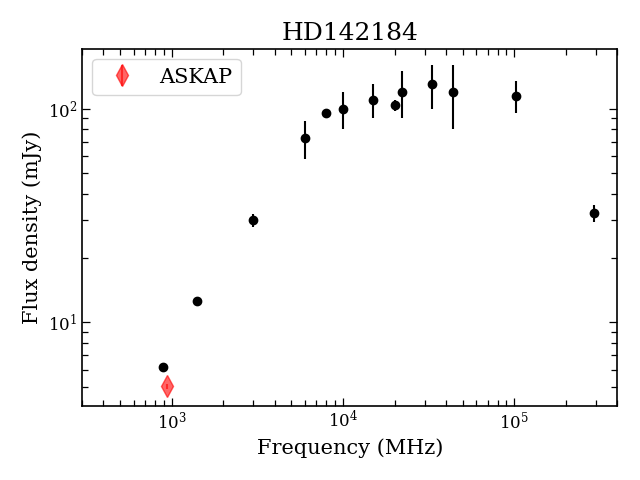}}
  \caption{Continued.}
\end{figure*}

\begin{figure*}
\ContinuedFloat 
  \centering 
  \subfloat{\includegraphics[width=0.45\textwidth]{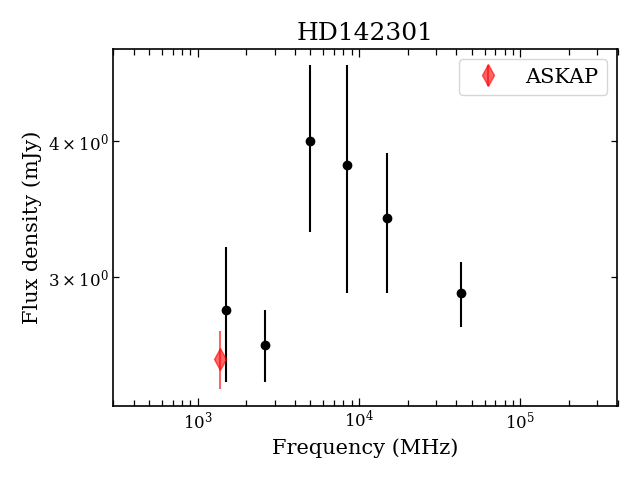}}
  \qquad 
  \subfloat{\includegraphics[width=0.45\textwidth]{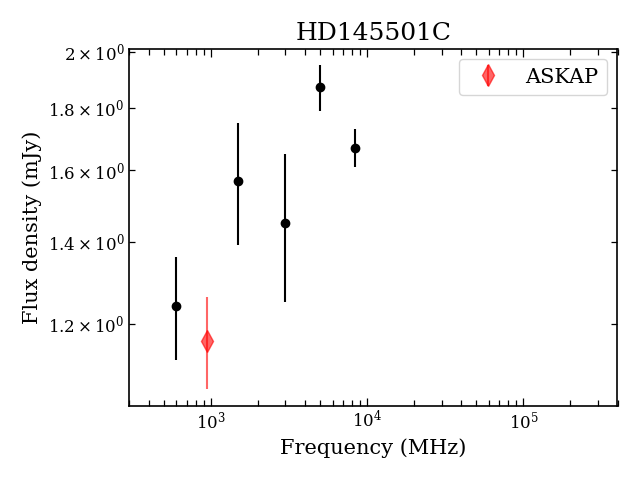}}
  \qquad 
  \subfloat{\includegraphics[width=0.45\textwidth]{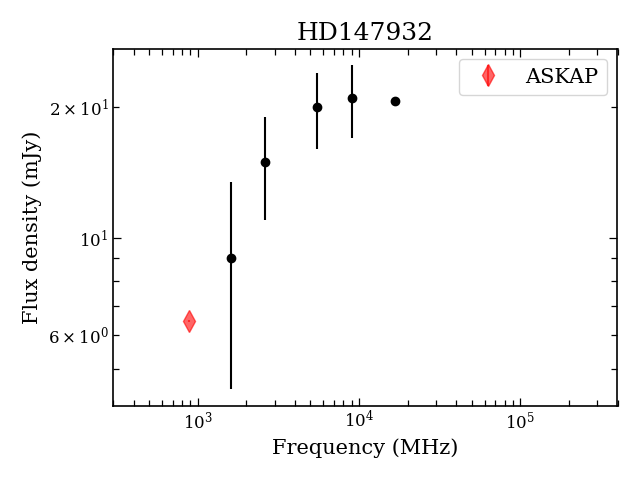}}
  \qquad 
  \subfloat{\includegraphics[width=0.45\textwidth]{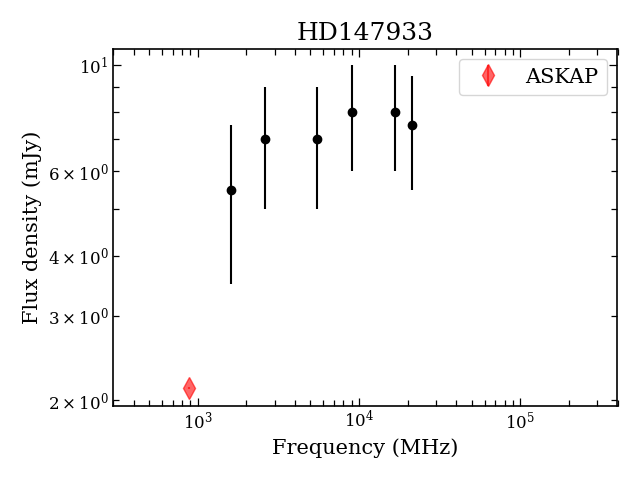}}
  \qquad 
  \subfloat{\includegraphics[width=0.45\textwidth]{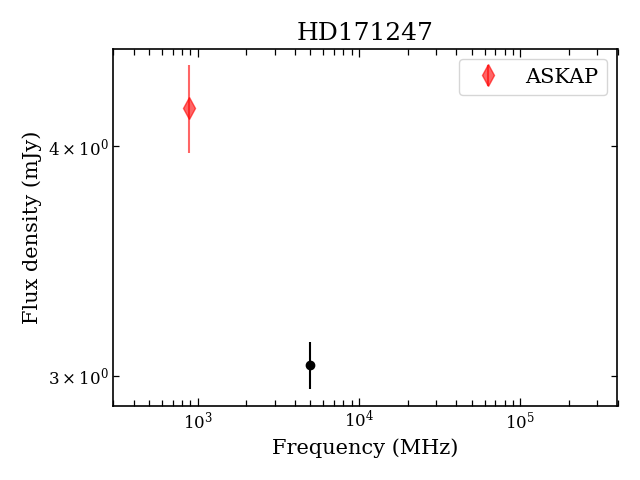}}
  \qquad 
  \subfloat{\includegraphics[width=0.45\textwidth]{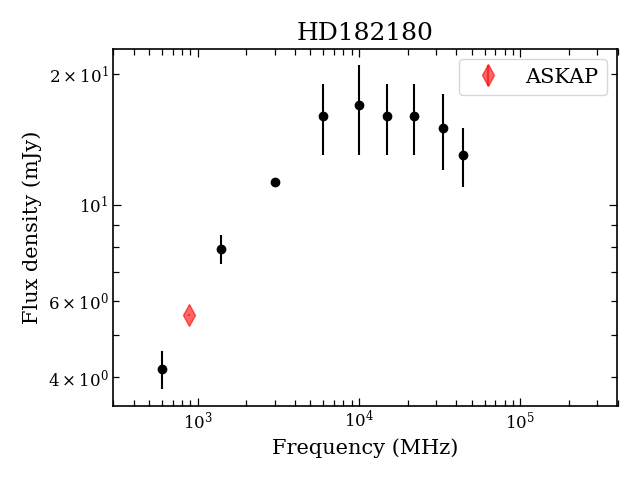}}
  \caption{Continued.}
\end{figure*}

\begin{table*}[]
    \centering
    \caption{Cross-match results for the short radio catalogue. The separations column shows the smallest separation between the proper motion corrected position of the star and a radio source. The radio component column gives the unique \textsc{Selavy} identifier for the ASKAP detection. The SB of the observation can be found in this identifier, which can be used to query CASDA to find the observation.}
\begin{tabular}{llllP{1.6cm}l}
\hline
Star & Radio component & Radio position & Obs. date & Integration time (s) & Separation \\
\hline
HD 12447 & SB33442\_component\_3523a & 02h02m02.88s$\pm$0\farcs4 +02d45m49.0s$\pm$0\farcs4 & 2021-11-10 & 18015 & 0\farcs6 \\
HD 22470 & SB63026\_component\_7408a & 03h36m17.56s$\pm$1\arcsec\, $-$17d28m02.2s$\pm$0\farcs6 & 2024-06-17 & 7216 & 1\farcs4 \\
HD 25267 & SB37453\_component\_12112a & 03h59m55.35s$\pm$1\arcsec\, $-$24d00m58.5s$\pm$0\farcs8 & 2022-02-18 & 7206 & 2\farcs2 \\
HD 37808 & SB62244\_component\_2507a & 05h40m46.15s$\pm$0\farcs2 $-$10d24m31.0s$\pm$0\farcs2 & 2024-05-09 & 18016 & 0\farcs7 \\
HD 56350 & SB51927\_component\_13125a & 07h13m40.04s$\pm$0\farcs7 $-$53d40m04.4s$\pm$0\farcs8 & 2023-08-09 & 36031 & 1\farcs7 \\
HD 64740 & SB54098\_component\_21806a & 07h53m03.59s$\pm$1\arcsec\, $-$49d36m48.6s$\pm$1\arcsec\, & 2023-10-21 & 36021 & 2\farcs0 \\
HD 83625 & SB51428\_component\_4226a & 09h38m03.13s$\pm$0\farcs3 $-$54d13m08.6s$\pm$0\farcs3 & 2023-07-13 & 36020 & 0\farcs5 \\
HD 101412 & SB60586\_component\_3865a & 11h39m44.79s$\pm$0\farcs3 $-$60d10m28.1s$\pm$0\farcs3 & 2024-04-01 & 35991 & 2\farcs6 \\
HD 105382 & SB64419\_component\_1899a & 12h08m05.230s$\pm$0\farcs1 $-$50d39m41.2s$\pm$0\farcs1 & 2024-08-08 & 36010 & 0\farcs9 \\
HD 112381 & SB45638\_component\_4974a & 12h56m58.16s$\pm$0\farcs3 $-$54d35m15.1s$\pm$0\farcs3 & 2022-11-16 & 36010 & 0\farcs4 \\
HD 118913 & SB50429\_component\_22413a & 13h42m26.93s$\pm$1\arcsec\, $-$69d14m46.2s$\pm$1\arcsec\, & 2023-06-11 & 36020 & 3\farcs0 \\
HD 122451 & SB54095\_component\_2526a & 14h03m49.23s$\pm$0\farcs4 $-$60d22m22.5s$\pm$0\farcs4 & 2023-10-21 & 36020 & 1\farcs1 \\
HD 124224 & SB14082\_component\_1278a & 14h12m15.754s$\pm$0\farcs03 +02d24m33.31s$\pm$0\farcs03 & 2020-05-06 & 31810 & 0\farcs1 \\
HD 128898 & SB45821\_component\_6668a & 14h42m29.72s$\pm$0\farcs4 $-$64d58m36.0s$\pm$0\farcs4 & 2022-11-24 & 36010 & 0\farcs2 \\
HD 142184 & SB50910\_component\_976a & 15h53m55.80s$\pm$0\farcs2 $-$23d58m41.3s$\pm$0\farcs2 & 2023-06-30 & 1214 & 0\farcs7 \\
HD 142301 & SB50910\_component\_618a & 15h54m39.480s$\pm$0\farcs1 $-$25d14m38.1s$\pm$0\farcs1 & 2023-06-30 & 1214 & 0\farcs4 \\
HD 145501C & SB52657\_component\_1222a & 16h11m58.465s$\pm$0\farcs07 $-$19d27m00.51s$\pm$0\farcs06 & 2023-09-07 & 28814 & 1\farcs2 \\
HD 148898 & SB52657\_component\_11588a & 16h32m08.12s$\pm$0\farcs6 $-$21d27m58.0s$\pm$0\farcs6 & 2023-09-07 & 28814 & 1\farcs5 \\
HD 149764 & SB28280\_component\_3520a & 16h38m30.89s$\pm$0\farcs2 $-$39d09m08.8s$\pm$0\farcs2 & 2021-06-26 & 36011 & 0\farcs3 \\
HD 151965 & SB28280\_component\_3334a & 16h52m27.37s$\pm$0\farcs2 $-$40d43m23.6s$\pm$0\farcs2 & 2021-06-26 & 36011 & 0\farcs4 \\
HD 170397 & SB48586\_component\_6213a & 18h29m46.83s$\pm$0\farcs7 $-$14d34m56.9s$\pm$0\farcs7 & 2023-02-24 & 21617 & 2\farcs1 \\
HD 221006 & SB53321\_component\_9976a & 23h29m01.59s$\pm$0\farcs6 $-$63d06m40.5s$\pm$0\farcs6 & 2023-10-01 & 36010 & 3\farcs7 \\
\hline
\end{tabular}
    \label{tab: short star results}
\end{table*}

\begin{table*}[]
    \centering
    \caption{Cross-match results for the long radio catalogue. The separations column shows the smallest separation between the proper motion corrected position of the star and a radio source. The radio component column gives the unique \textsc{Selavy} identifier for the ASKAP detection. The SB of the observation can be found in this identifier, which can be used to query CASDA to find the observation.}
\begin{tabular}{llllP{1.6cm}l}
\hline
Star & Radio component & Radio position & Obs. date & Integration time (s) & Separation \\
\hline
HD 315 & SB28894\_component\_4030a & 00h07m44.20s$\pm$0\farcs8 $-$02d32m54.1s$\pm$0\farcs7 & 2021-07-21 & 717 & 1\farcs3 \\
HD 12447 & SB56485\_component\_3825a & 02h02m02.89s$\pm$0\farcs8 +02d45m50.0s$\pm$0\farcs7 & 2024-01-01 & 835 & 0\farcs6 \\
HD 22470 & SB35733\_component\_2393a & 03h36m17.39s$\pm$0\farcs6 $-$17d28m01.0s$\pm$0\farcs5 & 2022-01-11 & 905 & 1\farcs1 \\
HD 23478 & SB20297\_component\_1331a & 03h46m40.78s$\pm$0\farcs3 +32d17m24.6s$\pm$0\farcs5 & 2020-12-24 & 905 & 1\farcs3 \\
HD 27309 & SB34949\_component\_2252a & 04h19m36.84s$\pm$0\farcs7 +21d46m20.4s$\pm$1\arcsec\, & 2021-12-29 & 895 & 3\farcs5 \\
HD 34736 & SB35608\_component\_1925a & 05h19m21.24s$\pm$0\farcs3 $-$07d20m50.7s$\pm$0\farcs3 & 2022-01-09 & 905 & 0\farcs7 \\
HD 35298 & SB51394\_component\_2608a & 05h23m50.33s$\pm$0\farcs5 +02d04m55.4s$\pm$0\farcs5 & 2023-07-08 & 726 & 0\farcs6 \\
HD 35502 & SB21057\_component\_1928a & 05h25m01.22s$\pm$0\farcs4 $-$02d48m54.6s$\pm$0\farcs3 & 2021-01-09 & 905 & 1\farcs1 \\
HD 36485 & SB54484\_component\_5151a & 05h32m00.39s$\pm$1\arcsec\, $-$00d17m04.5s$\pm$1\arcsec\, & 2023-10-30 & 727 & 0\farcs4 \\
HD 36526 & SB51007\_component\_4959a & 05h32m13.22s$\pm$0\farcs9 $-$01d36m03.5s$\pm$0\farcs8 & 2023-07-02 & 717 & 2\farcs7 \\
HD 37017 & SB35470\_component\_2044a & 05h35m21.87s$\pm$0\farcs7 $-$04d29m40.1s$\pm$0\farcs6 & 2022-01-07 & 896 & 1\farcs1 \\
HD 37808 & SB21654\_component\_1817a & 05h40m46.17s$\pm$0\farcs4 $-$10d24m30.8s$\pm$0\farcs3 & 2021-01-20 & 905 & 0\farcs5 \\
HD 40312 & SB20645\_component\_2661a & 05h59m43.27s$\pm$1\arcsec\, +37d12m43.1s$\pm$2\arcsec\, & 2021-01-01 & 905 & 1\farcs2 \\
HD 61556 & SB51166\_component\_5488a & 07h38m49.83s$\pm$0\farcs8 $-$26d48m12.8s$\pm$0\farcs7 & 2023-07-05 & 727 & 0\farcs5 \\
HD 83625 & SB61422\_component\_2647a & 09h38m03.14s$\pm$0\farcs3 $-$54d13m08.7s$\pm$0\farcs4 & 2024-04-21 & 726 & 0\farcs4 \\
HD 92385 & SB50129\_component\_1912a & 10h38m17.65s$\pm$0\farcs2 $-$65d02m31.3s$\pm$0\farcs2 & 2023-05-21 & 717 & 1\farcs2 \\
HD 101412 & SB22343\_component\_2866a & 11h39m44.83s$\pm$0\farcs6 $-$60d10m28.6s$\pm$0\farcs7 & 2021-02-01 & 905 & 3\farcs0 \\
HD 105382 & SB36395\_component\_876a & 12h08m05.220s$\pm$0\farcs1 $-$50d39m40.8s$\pm$0\farcs1 & 2022-01-23 & 904 & 0\farcs7 \\
HD 112413 & SB35149\_component\_1255a & 12h56m01.29s$\pm$0\farcs2 +38d19m06.8s$\pm$0\farcs5 & 2022-01-01 & 896 & 0\farcs9 \\
HD 121743 & SB59855\_component\_4180a & 13h58m16.27s$\pm$0\farcs8 $-$42d06m02.7s$\pm$0\farcs8 & 2024-03-07 & 727 & 0\farcs8 \\
HD 124224 & SB36223\_component\_1074a & 14h12m15.73s$\pm$0\farcs6 +02d24m33.0s$\pm$0\farcs4 & 2022-01-20 & 905 & 0\farcs3 \\
HD 133880 & SB36603\_component\_1527a & 15h08m12.08s$\pm$0\farcs3 $-$40d35m03.0s$\pm$0\farcs3 & 2022-01-27 & 904 & 0\farcs2 \\
HD 134759 & SB58150\_component\_3431a & 15h12m13.22s$\pm$0\farcs7 $-$19d47m31.1s$\pm$0\farcs6 & 2024-01-28 & 846 & 0\farcs3 \\
HD 136347 & SB22062\_component\_2967a & 15h21m30.08s$\pm$0\farcs6 $-$38d13m07.6s$\pm$0\farcs5 & 2021-01-26 & 905 & 0\farcs2 \\
HD 136504A & SB36490\_component\_1213a & 15h22m40.85s$\pm$0\farcs3 $-$44d41m23.2s$\pm$0\farcs2 & 2022-01-25 & 905 & 0\farcs3 \\
HD 136504B & SB36490\_component\_1213a & 15h22m40.85s$\pm$0\farcs3 $-$44d41m23.2s$\pm$0\farcs2 & 2022-01-25 & 905 & 0\farcs3 \\
HD 142184 & SB58237\_component\_1571a & 15h53m55.81s$\pm$0\farcs1 $-$23d58m42.1s$\pm$0\farcs1 & 2024-01-30 & 915 & 0\farcs5 \\
HD 142301 & SB36158\_component\_723a & 15h54m39.490s$\pm$0\farcs1 $-$25d14m38.40s$\pm$0\farcs09 & 2022-01-20 & 915 & 0\farcs5 \\
HD 142990 & SB36158\_component\_829a & 15h58m34.83s$\pm$0\farcs1 $-$24d49m53.9s$\pm$0\farcs1 & 2022-01-20 & 915 & 0\farcs2 \\
HD 145501C & SB35698\_component\_1688a & 16h11m58.49s$\pm$0\farcs3 $-$19d27m01.7s$\pm$0\farcs3 & 2022-01-11 & 905 & 0\farcs1 \\
HD 147932 & SB36159\_component\_309a & 16h25m35.055s$\pm$0\farcs05 $-$23d24m19.02s$\pm$0\farcs04 & 2022-01-20 & 896 & 0\farcs3 \\
HD 147933 & SB36159\_component\_487a & 16h25m35.104s$\pm$0\farcs09 $-$23d26m50.23s$\pm$0\farcs08 & 2022-01-20 & 896 & 0\farcs2 \\
HD 149764 & SB58448\_component\_1126a & 16h38m30.92s$\pm$0\farcs2 $-$39d09m09.1s$\pm$0\farcs1 & 2024-02-01 & 726 & 0\farcs2 \\
HD 151965 & SB21434\_component\_2384a & 16h52m27.40s$\pm$0\farcs4 $-$40d43m23.4s$\pm$0\farcs4 & 2021-01-17 & 915 & 0\farcs4 \\
HD 164224 & SB64111\_component\_2733a & 18h00m57.58s$\pm$0\farcs8 $-$20d59m17.8s$\pm$0\farcs7 & 2024-07-21 & 727 & 0\farcs6 \\
HD 170397 & SB21512\_component\_2443a & 18h29m46.82s$\pm$0\farcs6 $-$14d34m55.1s$\pm$0\farcs5 & 2021-01-18 & 894 & 0\farcs5 \\
HD 171247 & SB61216\_component\_2283a & 18h33m23.55s$\pm$0\farcs4 +08d16m09.6s$\pm$0\farcs4 & 2024-04-17 & 737 & 5\farcs6 \\
HD 182180 & SB55678\_component\_1096a & 19h24m30.218s$\pm$0\farcs1 $-$27d51m57.6s$\pm$0\farcs1 & 2023-12-22 & 905 & 0\farcs3 \\
HD 196178 & SB34702\_component\_1452a & 20h33m55.11s$\pm$0\farcs3 +46d41m33.2s$\pm$1\arcsec\, & 2021-12-25 & 915 & 5\farcs4 \\
\hline
\end{tabular}
    \label{tab: long star results}
\end{table*}

\begin{table*}[]
    \centering
    \caption{Cross-match results for the circular polarisation search. The separations column shows the smallest separation between the proper motion corrected position of the star and a radio source. The radio component column gives the unique \textsc{Selavy} identifier for the ASKAP detection. The SB of the observation can be found in this identifier, which can be used to query CASDA to find the observation. The boldfaced stars represent new MRP candidates, the others are known MRPs.}
\begin{tabular}{llllP{1.6cm}l}
\hline
Star & Radio component & Radio position & Obs. date & ${\left|V\right|}/{I}$ & Separation \\
\hline
\textbf{HD 34736} & SB35608\_component\_1925a & 05h19m21.24s$\pm$0\farcs3 $-$07d20m50.7s$\pm$0\farcs3 & 2022-01-09 & 0.664547 & 0\farcs7 \\
\textbf{HD 37808} & SB62244\_component\_2507a & 05h40m46.15s$\pm$0\farcs2 $-$10d24m31.0s$\pm$0\farcs2 & 2024-05-09 & 0.369867 & 0\farcs7 \\
HD 61556 & SB20583\_component\_1839a & 07h38m49.92s$\pm$0\farcs3 $-$26d48m13.3s$\pm$0\farcs3 & 2020-12-31 & 0.568918 & 1\farcs0 \\
\textbf{HD 105382} & SB59494\_component\_4603a & 12h08m05.04s$\pm$0\farcs6 $-$50d39m40.5s$\pm$0\farcs7 & 2024-02-26 & 0.595534 & 0\farcs99 \\
\textbf{HD 122451} & SB54095\_component\_2526a & 14h03m49.23s$\pm$0\farcs4 $-$60d22m22.5s$\pm$0\farcs4 & 2023-10-21 & 0.395385 & 1\farcs1 \\
HD 124224 & SB30876\_component\_1055a & 14h12m15.730s$\pm$0\farcs1 +02d24m33.9s$\pm$0\farcs1 & 2021-08-21 & 0.152935 & 0\farcs6 \\
HD 142301 & SB58237\_component\_1475a & 15h54m39.50s$\pm$0\farcs1 $-$25d14m38.7s$\pm$0\farcs1 & 2024-01-30 & 0.406277 & 0\farcs7 \\
\textbf{HD 164224} & SB22482\_component\_1453a & 18h00m57.54s$\pm$0\farcs3 $-$20d59m16.2s$\pm$0\farcs3 & 2021-02-06 & 0.768174 & 1\farcs7 \\
\textbf{HD 196178} & SB34702\_component\_1452a & 20h33m55.11s$\pm$0\farcs3 +46d41m33.2s$\pm$1\arcsec\, & 2021-12-25 & 0.497540 & 5\farcs4 \\
\hline
\end{tabular}
    \label{tab: stokes V star results}
\end{table*}

\begin{table*}[hbt!]
\begin{threeparttable}
\caption{he stellar magnetospheric parameters and radio luminosities of magnetic massive stars detected in ASKAP survey data. `-1' indicates absence of information. The quoted uncertainty in the estimates of incoherent radio luminosity only incorporates the uncertainty in the flux density measurement. The other sources of uncertainties are discussed in the main text (\S\ref{subsec:Lrad_estimation} and \S\ref{subsec:indifference_to_Teff}).
$L_\mathrm{CBO}$ is calculated using Equation \ref{eq:Lcbo_expression}.
}
\label{tab:radio_detections}
\begin{tabular}{lllllllll}
\toprule
\headrow Star & $M\,(M_\odot)$ & $R\,(R_\odot)$ & $T_\mathrm{eff}\,(\mathrm{kK})$ & $B\,\mathrm{(kG)}$ & $P_\mathrm{rot}$ (d) & $\log (L_\mathrm{CBO}/L_\odot)$ & $\log(L_\mathrm{rad,ASKAP}/L_\odot)$ & Note\\
\midrule 
HD 12447&$2.6 \pm 0.9$ & $2.7 \pm 0.2$ & $10.0 \pm 0.2$ & $2.5^{+0.7}_{-0.5}$ & $1.491$ & $1.4^{+0.34}_{-0.27}$ & $-7.34 \pm 0.04$ & \\
HD 19832&$1.8 \pm 1.1$ & $2.2 \pm 0.4$ & $12.8 \pm 0.4$ & $2.7^{+0.7}_{-0.7}$ & $0.7278$ & $1.86^{+0.46}_{-0.55}$ & $-6.29 \pm 0.05$ & \\
HD 22470&$3.0 \pm 1.8$ & $2.9 \pm 0.5$ & $13.8 \pm 0.3$ & $7.5^{+1.1}_{-1.3}$ & $1.9289$ & $2.27^{+0.41}_{-0.45}$ & $-6.65^{+0.06}_{-0.08}$ & New addition\\
HD 23478&$6.0 \pm 4.9$ & $3.2 \pm 1.1$ & $20.0 \pm 1.9$ & $10.3^{+2.0}_{-2.4}$ & $1.0498$ & $3.15^{+0.63}_{-0.89}$ & $-5.56^{+0.02}_{-0.01}$ &\\
HD 25267&$-1$ & $3.3 \pm 0.3$ & $12.0 \pm 0.4$ & $\approx1.0$ & $9.3897$ & $-1$ & $-7.13 \pm 0.04$ & New addition\\
HD 27309&$2.9 \pm 0.8$ & $2.3 \pm 0.1$ & $11.3 \pm 0.2$ & $3.6^{+2.0}_{-0.6}$ & $1.569$ & $1.39^{+0.47}_{-0.21}$ & $-6.8^{+0.08}_{-0.1}$ & \\
HD 34736&$4.1 \pm 1.8$ & $2.8 \pm 0.4$ & $13.0 \pm 0.5$ & $18.9^{+0.8}_{-0.8}$ & $1.2799885$ & $3.33^{+0.28}_{-0.28}$ & $-5.46 \pm 0.05$ & \\
HD 35298&$2.9 \pm 1.4$ & $2.1 \pm 0.4$ & $15.8 \pm 0.8$ & $11.2^{+0.9}_{-1.1}$ & $1.8546$ & $2.06^{+0.38}_{-0.41}$ & $-5.21 \pm 0.04$ & \\
HD 35502&$6.3 \pm 3.5$ & $2.9 \pm 0.4$ & $18.4 \pm 0.6$ & $7.3^{+0.5}_{-0.5}$ & $0.8538$ & $2.84^{+0.35}_{-0.34}$ & $-5.37^{+0.04}_{-0.05}$ & \\
HD 36485&$5.1 \pm 3.8$ & $3.0 \pm 0.9$ & $20.0 \pm 1.9$ & $8.9^{+0.2}_{-0.3}$ & $1.4777$ & $2.6^{+0.55}_{-0.7}$ & $-5.49^{+0.07}_{-0.1}$ & \\
HD 36526&$2.0 \pm 1.6$ & $2.1 \pm 0.8$ & $15.0 \pm 1.9$ & $11.2^{+0.4}_{-0.6}$ & $1.5415$ & $2.29^{+0.65}_{-0.89}$ & $-5.41^{+0.05}_{-0.06}$ & New addition\\
HD 37017&$6.9 \pm 5.7$ & $3.9 \pm 1.3$ & $21.0 \pm 1.9$ & $6.2^{+0.8}_{-1.0}$ & $0.9012$ & $3.18^{+0.63}_{-0.83}$ & $-5.51^{+0.09}_{-0.12}$ & \\
HD 37479&$7.4 \pm 5.6$ & $3.6 \pm 1.1$ & $23.0 \pm 1.9$ & $10.5^{+1.5}_{-1.5}$ & $1.1908$ & $3.22^{+0.56}_{-0.69}$ & $-5.24^{+0.07}_{-0.08}$ & \\
HD 37808&$2.2 \pm 1.6$ & $2.2 \pm 0.3$ & $14.5 \pm 0.5$ & $3.2^{+0.4}_{-0.4}$ & $1.0991$ & $1.56^{+0.34}_{-0.33}$ & $-6.08^{+0.07}_{-0.09}$ & \\
HD 40312&$3.3 \pm 0.6$ & $4.7 \pm 0.1$ & $10.2 \pm 0.1$ & $0.9^{+2.8}_{-0.1}$ & $3.619$ & $0.8^{+1.37}_{-0.24}$ & $-7.28^{+0.08}_{-0.09}$ & \\
HD 56350&$-1$ & $2.4 \pm 0.3$ & $10.5 \pm 0.5$ & $\approx3.1$ & $-1$ & $-1$ & $-7.02^{+0.04}_{-0.05}$ & New addition\\
HD 61556&$5.8 \pm 3.9$ & $3.5 \pm 1.0$ & $18.5 \pm 0.8$ & $2.8^{+0.3}_{-0.3}$ & $1.9087$ & $1.71^{+0.55}_{-0.66}$ & $-6.57 \pm 0.03$ & \\
HD 64740&$7.5 \pm 3.2$ & $4.5 \pm 0.8$ & $24.5 \pm 1.0$ & $3.0^{+0.4}_{-0.5}$ & $1.3302$ & $2.48^{+0.4}_{-0.45}$ & $-6.72^{+0.07}_{-0.08}$ & \\
HD 83625&$3.0 \pm 1.7$ & $2.1 \pm 0.3$ & $11.7 \pm 0.5$ & $\approx4.4$ & $1.0784747$ & $1.7^{+0.0}_{-0.0}$ & $-6.39 \pm 0.02$ & \\
HD 92385&$-1$ & $2.2 \pm 0.3$ & $11.2 \pm 0.5$ & $\approx1.5$ & $0.549$ & $-1$ & $-5.78 \pm 0.02$ & New addition\\
HD 101412&$6.2 \pm 7.6$ & $4.1 \pm 0.9$ & $8.6 \pm 0.6$ & $1.8^{+0.2}_{-0.2}$ & $42.076$ & $-1.12^{+0.43}_{-0.56}$ & $-5.58^{+0.02}_{-0.03}$ & New addition\\
HD 105382&$5.8 \pm 2.4$ & $3.4 \pm 0.7$ & $18.0 \pm 0.5$ & $2.6^{+0.1}_{-0.1}$ & $1.2950709$ & $1.88^{+0.39}_{-0.4}$ & $-6.62^{+0.05}_{-0.06}$ & \\
HD 112381&$2.2 \pm 0.5$ & $1.7 \pm 0.2$ & $10.0 \pm 0.3$ & $\approx11.9$ & $2.84$ & $1.33^{+0.0}_{-0.0}$ & $-6.75 \pm 0.03$ & \\
HD 112413&$2.0 \pm 0.5$ & $2.9 \pm 0.1$ & $11.3 \pm 0.2$ & $4.2^{+2.3}_{-0.7}$ & $5.469$ & $1.0^{+0.47}_{-0.21}$ & $-7.72^{+0.08}_{-0.1}$ & \\
HD 118913&$-1$ & $2.5 \pm 0.4$ & $9.6 \pm 0.2$ & $\approx1.4$ & $-1$ & $-1$ & $-6.93^{+0.22}_{-0.49}$ & New addition\\
HD 121743&$9.6 \pm 5.1$ & $5.0 \pm 1.1$ & $21.0 \pm 1.3$ & $1.1^{+0.3}_{-0.3}$ & $1.1302$ & $1.91^{+0.52}_{-0.62}$ & $-6.25^{+0.06}_{-0.07}$ & \\
HD 122451&$13.6 \pm 8.7$ & $10.2 \pm 3.0$ & $23.0 \pm 1.9$ & $0.2^{+0.0}_{-0.0}$ & $2.8849$ & $1.03^{+0.58}_{-0.72}$ & $-6.34^{+0.07}_{-0.08}$ & New addition\\
HD 124224&$3.1 \pm 0.4$ & $2.1 \pm 0.1$ & $12.2 \pm 0.2$ & $2.3^{+0.8}_{-0.6}$ & $0.5207$ & $1.71^{+0.35}_{-0.32}$ & $-6.7^{+0.05}_{-0.06}$ & \\
HD 128898&$1.8 \pm 0.3$ & $1.9 \pm 0.0$ & $7.8 \pm 0.1$ & $1.4^{+0.3}_{-0.3}$ & $4.479$ & $-0.64^{+0.22}_{-0.21}$ & $-8.63 \pm 0.03$ & \\
HD 133880&$-1$ & $2.6 \pm 1.4$ & $12.0 \pm 2.9$ & $\approx8.0$ & $0.877$ & $-1$ & $-6.43^{+0.03}_{-0.04}$ & \\
HD 134759&$3.6 \pm 1.1$ & $4.0 \pm 0.4$ & $11.9 \pm 0.2$ & $\approx1.3$ & $-1$ & $-1$ & $-6.41^{+0.06}_{-0.07}$ & New addition\\
HD 136347&$1.9 \pm 0.6$ & $2.0 \pm 0.3$ & $11.4 \pm 0.7$ & $\approx0.8$ & $-1$ & $-1$ & $-6.42^{+0.05}_{-0.07}$ & New addition\\
HD 136504B&$10.5 \pm 6.2$ & $4.6 \pm 1.1$ & $18.5 \pm 0.5$ & $0.5^{+0.2}_{-0.2}$ & $-1$ & $-1$ & $-5.93^{+0.09}_{-0.11}$ & \\
HD 142184&$5.0 \pm 1.7$ & $2.6 \pm 0.4$ & $18.5 \pm 0.5$ & $9.0^{+2.7}_{-2.3}$ & $0.5083$ & $3.28^{+0.45}_{-0.45}$ & $-5.69^{+0.02}_{-0.01}$ & \\
HD 142301&$2.3 \pm 1.1$ & $2.5 \pm 0.2$ & $15.9 \pm 0.2$ & $13.4^{+2.3}_{-2.3}$ & $1.4594$ & $2.82^{+0.3}_{-0.27}$ & $-5.98 \pm 0.03$ & \\
HD 145501C&$2.6 \pm 1.5$ & $2.7 \pm 0.5$ & $14.5 \pm 0.5$ & $5.8^{+0.4}_{-0.5}$ & $1.0265$ & $2.49^{+0.39}_{-0.41}$ & $-6.33^{+0.03}_{-0.04}$ & \\
HD 147932&$1.2 \pm 1.1$ & $1.8 \pm 0.7$ & $17.0 \pm 1.0$ & $7.6^{+9.6}_{-0.5}$ & $0.8638971$ & $2.3^{+1.08}_{-0.73}$ & $-5.71^{+0.01}_{-0.0}$ & \\
HD 147933&$5.8 \pm 4.6$ & $3.0 \pm 0.8$ & $19.1 \pm 2.2$ & $4.0^{+2.8}_{-0.2}$ & $0.747326$ & $2.48^{+0.79}_{-0.56}$ & $-6.09^{+0.01}_{-0.02}$ & \\
HD 148898&$2.3 \pm 0.3$ & $3.0 \pm 0.1$ & $8.1 \pm 0.1$ & $2.6^{+0.4}_{-0.3}$ & $2.321$ & $1.31^{+0.18}_{-0.15}$ & $-7.98^{+0.07}_{-0.08}$ & New addition\\
HD 148937&$90.8 \pm 34.4$ & $15.8 \pm 2.4$ & $41.0 \pm 2.0$ & $\approx1.0$ & $7.0323$ & $2.0^{+0.0}_{-0.0}$ & $-4.07 \pm 0.06$ & \\
HD 149764&$3.2 \pm 1.8$ & $2.1 \pm 0.3$ & $12.9 \pm 0.6$ & $\approx2.9$ & $0.63934468$ & $1.77^{+0.0}_{-0.0}$ & $-6.2^{+0.06}_{-0.07}$ & \\
HD 151965&$-1$ & $2.4 \pm 0.4$ & $13.8 \pm 0.5$ & $\approx26.7$ & $1.60866$ & $-1$ & $-6.09 \pm 0.03$ & New addition\\
HD 164224&$0.5 \pm 0.1$ & $2.2 \pm 0.1$ & $8.9 \pm 0.2$ & $\approx2.0$ & $0.73$ & $1.82^{+0.0}_{-0.0}$ & $-5.8^{+0.09}_{-0.11}$ & \\
HD 170397&$-1$ & $2.1 \pm 0.3$ & $10.0 \pm 0.5$ & $\approx3.8$ & $2.2545$ & $-1$ & $-6.57 \pm 0.05$ & New addition\\
HD 171247&$-1$ & $8.9 \pm 2.5$ & $11.7 \pm 0.5$ & $\approx4.6$ & $3.912$ & $-1$ & $-5.03 \pm 0.03$ & \\
HD 182180&$5.8 \pm 4.6$ & $3.0 \pm 1.2$ & $20.0 \pm 3.2$ & $9.5^{+0.6}_{-0.6}$ & $0.5214$ & $3.54^{+0.71}_{-0.92}$ & $-5.25 \pm 0.01$ & \\
HD 221006&$-1$ & $2.2 \pm 0.3$ & $13.4 \pm 0.5$ & $\approx3.5$ & $2.315$ & $-1$ & $-7.09 \pm 0.04$ & New addition\\
\bottomrule
\end{tabular}
\end{threeparttable}
\end{table*}

\newpage
\clearpage
\onecolumn
\begin{center}
\begin{longtable}{lrccr}
\caption{Stokes I Flux densities of magnetic hot stars measured by ASKAP at different epochs of observations.}\label{tab:radio_fluxes}\\
\hline
\textbf{Star} & \textbf{Frequency (MHz)} & \textbf{Start time (MJD)} & \textbf{End time (MJD)} & \textbf{Flux density (mJy)} \\
\hline 
\endfirsthead

\hline
\textbf{Star} & \textbf{Frequency (MHz)} & \textbf{Start time (MJD)} & \textbf{End time (MJD)} & \textbf{Flux density (mJy)} \\
\hline
\endhead

\hline
\multicolumn{3}{r}{\emph{Continued on next page}} \\
\endfoot

\hline
\endlastfoot
HD 12447&$887.5$ & $60130.01375$ & $60130.02216$ & $6.29 \pm 0.43$\\
HD 12447&$887.5$ & $60664.50209$ & $60664.51038$ & $1.58 \pm 0.31$\\
HD 12447&$887.5$ & $58784.58447$ & $58784.83745$ & $1.48 \pm 0.03$\\
HD 12447&$855.5$ & $59566.62595$ & $59566.70935$ & $2.49 \pm 0.1$\\
HD 12447&$943.5$ & $59528.59822$ & $59528.80672$ & $1.45 \pm 0.08$\\
HD 12447&$887.5$ & $59416.92866$ & $59416.93707$ & $3.59 \pm 0.34$\\
HD 12447&$887.5$ & $59019.94758$ & $59019.956$ & $1.7 \pm 0.02$\\
HD 12447&$943.5$ & $59523.49793$ & $59523.70644$ & $1.12 \pm 0.09$\\
HD 12447&$887.5$ & $58866.39615$ & $58866.40455$ & $2.91 \pm 0.02$\\
HD 12447&$887.5$ & $60367.37507$ & $60367.38347$ & $3.57 \pm 0.84$\\
HD 12447&$887.5$ & $60304.45184$ & $60304.45944$ & $2.49 \pm 0.42$\\
HD 12447&$887.5$ & $60246.59432$ & $60246.60273$ & $2.67 \pm 0.24$\\
HD 12447&$855.5$ & $60141.01839$ & $60141.10191$ & $0.9 \pm 0.06$\\
HD 12447&$887.5$ & $60130.04209$ & $60130.0505$ & $7.5 \pm 0.64$\\
HD 12447&$887.5$ & $58867.39337$ & $58867.40178$ & $2.01 \pm 0.01$\\
HD 12447&$943.5$ & $60310.51207$ & $60310.52175$ & $1.13 \pm 0.13$\\
HD 19832&$943.5$ & $60305.58365$ & $60305.59413$ & $1.58 \pm 0.18$\\
HD 22470&$1655.5$ & $59590.55263$ & $59590.56311$ & $0.99 \pm 0.18$\\
HD 22470&$855.5$ & $60478.18595$ & $60478.26947$ & $0.8 \pm 0.09$\\
HD 23478&$1367.5$ & $59207.5478$ & $59207.55829$ & $3.32 \pm 0.22$\\
HD 23478&$1655.5$ & $59575.6101$ & $59575.62071$ & $2.83 \pm 0.21$\\
HD 23478&$887.5$ & $58594.29209$ & $58594.41973$ & $2.18 \pm 0.01$\\
HD 25267&$943.5$ & $60313.57983$ & $60313.5902$ & $1.04 \pm 0.1$\\
HD 25267&$855.5$ & $59628.52016$ & $59628.60356$ & $0.46 \pm 0.07$\\
HD 25267&$1367.5$ & $60660.47741$ & $60660.8109$ & $0.37 \pm 0.03$\\
HD 27309&$1655.5$ & $59577.61419$ & $59577.62455$ & $1.03 \pm 0.22$\\
HD 34736&$887.5$ & $60133.08154$ & $60133.08994$ & $1.25 \pm 0.14$\\
HD 34736&$887.5$ & $60247.77581$ & $60247.78422$ & $3.08 \pm 0.2$\\
HD 34736&$887.5$ & $60663.65372$ & $60663.66225$ & $3.36 \pm 0.19$\\
HD 34736&$887.5$ & $60425.24995$ & $60425.25836$ & $1.79 \pm 0.2$\\
HD 35298&$887.5$ & $60133.1101$ & $60133.11852$ & $2.41 \pm 0.2$\\
HD 35502&$887.5$ & $60127.17003$ & $60127.17833$ & $1.76 \pm 0.22$\\
HD 35502&$1655.5$ & $59590.62302$ & $59590.63361$ & $2.04 \pm 0.19$\\
HD 35502&$1367.5$ & $59223.62546$ & $59223.63595$ & $1.6 \pm 0.15$\\
HD 36485&$887.5$ & $60247.78549$ & $60247.7939$ & $1.1 \pm 0.21$\\
HD 36485&$887.5$ & $60597.8285$ & $60597.83691$ & $1.28 \pm 0.2$\\
HD 36526&$887.5$ & $60127.17003$ & $60127.17833$ & $1.1 \pm 0.14$\\
HD 37017&$1655.5$ & $59586.64684$ & $59586.65721$ & $1.2 \pm 0.28$\\
HD 37479&$943.5$ & $60309.66336$ & $60309.67373$ & $1.54 \pm 0.24$\\
HD 37479&$1655.5$ & $59586.64684$ & $59586.65721$ & $2.17 \pm 0.24$\\
HD 37479&$887.5$ & $60597.8285$ & $60597.83691$ & $1.65 \pm 0.24$\\
HD 37479&$1367.5$ & $59223.63883$ & $59223.6492$ & $1.84 \pm 0.21$\\
HD 37808&$943.5$ & $60313.63973$ & $60313.64941$ & $1.2 \pm 0.21$\\
HD 37808&$1655.5$ & $59588.64648$ & $59588.65696$ & $1.72 \pm 0.23$\\
HD 40312&$943.5$ & $60300.69726$ & $60300.70774$ & $1.02 \pm 0.2$\\
HD 40312&$1367.5$ & $59215.69154$ & $59215.70203$ & $1.1 \pm 0.25$\\
HD 40312&$887.5$ & $58594.29209$ & $58594.41973$ & $1.17 \pm 0.01$\\
HD 56350&$943.5$ & $60165.91475$ & $60166.33177$ & $0.23 \pm 0.02$\\
HD 56350&$943.5$ & $60181.87484$ & $60182.29163$ & $0.18 \pm 0.02$\\
HD 61556&$887.5$ & $59964.61645$ & $59964.62485$ & $1.12 \pm 0.14$\\
HD 61556&$887.5$ & $59978.57829$ & $59978.5867$ & $1.22 \pm 0.19$\\
HD 61556&$887.5$ & $59979.64819$ & $59979.65661$ & $1.1 \pm 0.17$\\
HD 61556&$887.5$ & $59998.51648$ & $59998.5249$ & $1.2 \pm 0.21$\\
HD 61556&$887.5$ & $60009.48391$ & $60009.49231$ & $1.13 \pm 0.19$\\
HD 61556&$887.5$ & $60040.47586$ & $60040.48427$ & $1.12 \pm 0.16$\\
HD 61556&$887.5$ & $60099.23684$ & $60099.24514$ & $1.86 \pm 0.18$\\
HD 61556&$887.5$ & $60120.18807$ & $60120.19635$ & $1.34 \pm 0.14$\\
HD 61556&$887.5$ & $60130.15579$ & $60130.1642$ & $1.04 \pm 0.15$\\
HD 61556&$887.5$ & $60265.86362$ & $60265.87204$ & $1.22 \pm 0.08$\\
HD 61556&$887.5$ & $60286.81808$ & $60286.8266$ & $1.14 \pm 0.12$\\
HD 61556&$887.5$ & $60303.66767$ & $60303.67609$ & $1.59 \pm 0.2$\\
HD 61556&$943.5$ & $60313.73062$ & $60313.741$ & $0.92 \pm 0.05$\\
HD 61556&$887.5$ & $60340.57329$ & $60340.58182$ & $1.02 \pm 0.24$\\
HD 61556&$887.5$ & $60359.62603$ & $60359.63444$ & $1.1 \pm 0.15$\\
HD 61556&$1655.5$ & $59588.73012$ & $59588.7406$ & $1.52 \pm 0.16$\\
HD 61556&$887.5$ & $58599.28138$ & $58599.53954$ & $2.9 \pm 0.02$\\
HD 64740&$943.5$ & $60238.75588$ & $60239.17279$ & $0.16 \pm 0.03$\\
HD 83625&$887.5$ & $59933.87964$ & $59933.88793$ & $1.05 \pm 0.12$\\
HD 83625&$887.5$ & $60421.51978$ & $60421.52819$ & $3.48 \pm 0.21$\\
HD 83625&$943.5$ & $60138.11641$ & $60138.53332$ & $0.68 \pm 0.03$\\
HD 83625&$887.5$ & $60098.41916$ & $60098.42744$ & $0.98 \pm 0.17$\\
HD 83625&$887.5$ & $58972.48713$ & $58972.49762$ & $1.09 \pm 0.01$\\
HD 92385&$887.5$ & $59897.9427$ & $59897.95111$ & $6.52 \pm 0.17$\\
HD 92385&$887.5$ & $60456.43376$ & $60456.44218$ & $6.29 \pm 0.17$\\
HD 92385&$887.5$ & $60557.10219$ & $60557.11049$ & $6.29 \pm 0.24$\\
HD 92385&$887.5$ & $60540.23856$ & $60540.24698$ & $6.23 \pm 0.18$\\
HD 92385&$887.5$ & $60519.20662$ & $60519.21502$ & $5.83 \pm 0.14$\\
HD 92385&$887.5$ & $60504.25117$ & $60504.25958$ & $6.66 \pm 0.22$\\
HD 92385&$887.5$ & $58610.35683$ & $58610.63573$ & $5.89 \pm 0.05$\\
HD 92385&$887.5$ & $58610.35683$ & $58610.63573$ & $6.01 \pm 0.03$\\
HD 92385&$1655.5$ & $59603.81914$ & $59603.8302$ & $3.57 \pm 0.16$\\
HD 92385&$887.5$ & $60485.32205$ & $60485.33046$ & $6.21 \pm 0.15$\\
HD 92385&$887.5$ & $60470.44316$ & $60470.45156$ & $6.37 \pm 0.24$\\
HD 92385&$887.5$ & $60440.51866$ & $60440.52707$ & $6.34 \pm 0.2$\\
HD 92385&$887.5$ & $60319.8523$ & $60319.8606$ & $6.99 \pm 0.28$\\
HD 92385&$887.5$ & $60439.4289$ & $60439.43743$ & $6.06 \pm 0.21$\\
HD 92385&$887.5$ & $60424.57674$ & $60424.58515$ & $5.32 \pm 0.24$\\
HD 92385&$887.5$ & $60410.56477$ & $60410.57318$ & $6.52 \pm 0.18$\\
HD 92385&$887.5$ & $60395.63494$ & $60395.64255$ & $6.04 \pm 0.18$\\
HD 92385&$887.5$ & $60357.72814$ & $60357.73655$ & $6.84 \pm 0.22$\\
HD 92385&$887.5$ & $59902.89507$ & $59902.90347$ & $6.42 \pm 0.16$\\
HD 92385&$943.5$ & $60321.81499$ & $60321.82536$ & $6.08 \pm 0.17$\\
HD 92385&$887.5$ & $60302.90378$ & $60302.91208$ & $7.12 \pm 0.68$\\
HD 92385&$887.5$ & $60198.17083$ & $60198.17925$ & $6.28 \pm 0.17$\\
HD 92385&$887.5$ & $60165.17925$ & $60165.18765$ & $5.76 \pm 0.17$\\
HD 92385&$887.5$ & $60184.16816$ & $60184.17656$ & $6.03 \pm 0.18$\\
HD 92385&$887.5$ & $60025.64669$ & $60025.65498$ & $5.66 \pm 0.16$\\
HD 92385&$887.5$ & $60055.57332$ & $60055.58162$ & $6.59 \pm 0.18$\\
HD 92385&$887.5$ & $60071.43856$ & $60071.44686$ & $6.1 \pm 0.18$\\
HD 92385&$887.5$ & $60085.48639$ & $60085.49469$ & $5.88 \pm 0.18$\\
HD 92385&$887.5$ & $60097.47166$ & $60097.48007$ & $5.98 \pm 0.16$\\
HD 92385&$887.5$ & $60129.36299$ & $60129.3714$ & $6.16 \pm 0.18$\\
HD 92385&$887.5$ & $60117.31208$ & $60117.32037$ & $6.16 \pm 0.18$\\
HD 92385&$887.5$ & $60118.40891$ & $60118.41721$ & $6.65 \pm 0.18$\\
HD 92385&$887.5$ & $60040.58196$ & $60040.59037$ & $5.98 \pm 0.22$\\
HD 92385&$887.5$ & $60101.35117$ & $60101.35958$ & $6.66 \pm 0.18$\\
HD 92385&$887.5$ & $59979.73494$ & $59979.74324$ & $6.49 \pm 0.19$\\
HD 92385&$887.5$ & $60009.60441$ & $60009.61271$ & $6.36 \pm 0.25$\\
HD 92385&$887.5$ & $59998.63525$ & $59998.64355$ & $6.55 \pm 0.22$\\
HD 92385&$887.5$ & $59993.65852$ & $59993.66681$ & $6.12 \pm 0.21$\\
HD 92385&$887.5$ & $60146.36931$ & $60146.37772$ & $6.34 \pm 0.2$\\
HD 92385&$887.5$ & $59978.6989$ & $59978.70731$ & $6.19 \pm 0.19$\\
HD 92385&$887.5$ & $59964.74546$ & $59964.75388$ & $6.11 \pm 0.18$\\
HD 92385&$887.5$ & $59949.85312$ & $59949.86153$ & $6.64 \pm 0.24$\\
HD 92385&$887.5$ & $59948.85676$ & $59948.86517$ & $6.48 \pm 0.19$\\
HD 92385&$887.5$ & $59934.81197$ & $59934.82037$ & $6.36 \pm 0.18$\\
HD 92385&$887.5$ & $59919.8524$ & $59919.86069$ & $6.31 \pm 0.18$\\
HD 92385&$887.5$ & $60265.92961$ & $60265.93812$ & $6.02 \pm 0.19$\\
HD 92385&$887.5$ & $60339.76206$ & $60339.77047$ & $5.44 \pm 0.16$\\
HD 101412&$887.5$ & $59902.022$ & $59902.03029$ & $1.22 \pm 0.16$\\
HD 101412&$887.5$ & $60100.49772$ & $60100.50625$ & $1.43 \pm 0.18$\\
HD 101412&$887.5$ & $60574.18705$ & $60574.19534$ & $1.21 \pm 0.18$\\
HD 101412&$1367.5$ & $59246.77551$ & $59246.78598$ & $0.92 \pm 0.14$\\
HD 101412&$943.5$ & $60401.48045$ & $60401.89701$ & $0.77 \pm 0.04$\\
HD 101412&$887.5$ & $60198.19274$ & $60198.20116$ & $1.11 \pm 0.17$\\
HD 101412&$887.5$ & $60183.27601$ & $60183.28441$ & $1.25 \pm 0.37$\\
HD 101412&$887.5$ & $60410.57468$ & $60410.58297$ & $1.29 \pm 0.24$\\
HD 105382&$887.5$ & $60120.48125$ & $60120.48965$ & $1.01 \pm 0.12$\\
HD 105382&$887.5$ & $60188.18426$ & $60188.19266$ & $1.22 \pm 0.18$\\
HD 105382&$887.5$ & $60245.03802$ & $60245.04631$ & $1.36 \pm 0.14$\\
HD 105382&$887.5$ & $60479.47257$ & $60479.48097$ & $1.41 \pm 0.13$\\
HD 105382&$887.5$ & $58609.36557$ & $58609.62293$ & $2.11 \pm 0.04$\\
HD 105382&$887.5$ & $60544.28204$ & $60544.29056$ & $1.19 \pm 0.13$\\
HD 105382&$887.5$ & $60608.03312$ & $60608.04142$ & $1.76 \pm 0.18$\\
HD 105382&$887.5$ & $60666.98$ & $60666.9883$ & $1.27 \pm 0.16$\\
HD 112381&$943.5$ & $59899.9013$ & $59900.31809$ & $0.58 \pm 0.03$\\
HD 112413&$943.5$ & $60300.98181$ & $60300.99216$ & $1.01 \pm 0.2$\\
HD 112413&$1367.5$ & $59206.98647$ & $59206.99683$ & $1.26 \pm 0.14$\\
HD 112413&$1655.5$ & $59580.97045$ & $59580.98082$ & $2.62 \pm 0.18$\\
HD 118913&$943.5$ & $60106.40267$ & $60106.81958$ & $0.16 \pm 0.02$\\
HD 118913&$943.5$ & $60669.68902$ & $60670.10593$ & $0.13 \pm 0.08$\\
HD 121743&$887.5$ & $60118.55877$ & $60118.56706$ & $1.74 \pm 0.21$\\
HD 121743&$1367.5$ & $59240.94274$ & $59240.95311$ & $1.4 \pm 0.18$\\
HD 121743&$887.5$ & $60376.77019$ & $60376.7786$ & $1.73 \pm 0.25$\\
HD 122451&$943.5$ & $60336.89374$ & $60336.90422$ & $1.57 \pm 0.17$\\
HD 124224&$887.5$ & $60108.59757$ & $60108.60598$ & $2.89 \pm 0.24$\\
HD 124224&$855.5$ & $59729.44856$ & $59729.53208$ & $6.44 \pm 0.13$\\
HD 124224&$887.5$ & $60670.94012$ & $60670.94863$ & $2.43 \pm 0.15$\\
HD 124224&$887.5$ & $60668.04991$ & $60668.05819$ & $8.66 \pm 0.92$\\
HD 124224&$887.5$ & $60611.09852$ & $60611.10635$ & $4.19 \pm 0.26$\\
HD 124224&$887.5$ & $60608.10939$ & $60608.1178$ & $7.14 \pm 0.63$\\
HD 124224&$887.5$ & $58723.01611$ & $58723.23027$ & $2.49 \pm 0.05$\\
HD 124224&$887.5$ & $58785.91825$ & $58786.0065$ & $2.7 \pm 0.01$\\
HD 124224&$1364.5$ & $58975.48196$ & $58975.85014$ & $3.42 \pm 0.03$\\
HD 124224&$1655.5$ & $59599.97316$ & $59599.98363$ & $2.98 \pm 0.5$\\
HD 124224&$855.5$ & $59728.49471$ & $59728.57823$ & $3.9 \pm 0.09$\\
HD 124224&$1655.5$ & $59591.99016$ & $59592.00064$ & $2.99 \pm 0.35$\\
HD 124224&$855.5$ & $59566.10594$ & $59566.18934$ & $2.57 \pm 0.09$\\
HD 124224&$887.5$ & $59447.31072$ & $59447.31902$ & $11.08 \pm 0.21$\\
HD 124224&$842.5$ & $60126.66427$ & $60126.67833$ & $2.21 \pm 0.18$\\
HD 124224&$1367.5$ & $59239.88929$ & $59239.89966$ & $2.28 \pm 0.2$\\
HD 124224&$1364.5$ & $58974.48502$ & $58974.85285$ & $3.71 \pm 0.03$\\
HD 124224&$887.5$ & $60131.54565$ & $60131.55405$ & $1.73 \pm 0.22$\\
HD 124224&$943.5$ & $60377.65354$ & $60377.86205$ & $2.83 \pm 0.06$\\
HD 124224&$855.5$ & $58958.60544$ & $58958.85554$ & $2.98 \pm 0.03$\\
HD 124224&$887.5$ & $58874.90037$ & $58874.90878$ & $2.17 \pm 0.02$\\
HD 124224&$887.5$ & $58873.90308$ & $58873.91149$ & $2.8 \pm 0.05$\\
HD 124224&$887.5$ & $58864.91697$ & $58864.92538$ & $2.54 \pm 0.03$\\
HD 124224&$943.5$ & $60410.59534$ & $60410.80384$ & $2.47 \pm 0.06$\\
HD 124224&$943.5$ & $60314.99059$ & $60315.00095$ & $7.78 \pm 0.71$\\
HD 124224&$943.5$ & $60328.90728$ & $60328.91765$ & $1.95 \pm 0.34$\\
HD 124224&$943.5$ & $60328.89611$ & $60328.90648$ & $2.4 \pm 0.31$\\
HD 124224&$855.5$ & $58960.63792$ & $58960.72132$ & $6.15 \pm 0.02$\\
HD 124224&$887.5$ & $60305.05656$ & $60305.06498$ & $4.75 \pm 0.29$\\
HD 124224&$887.5$ & $60249.10852$ & $60249.11692$ & $2.9 \pm 0.26$\\
HD 124224&$887.5$ & $60188.25649$ & $60188.26478$ & $6.42 \pm 0.24$\\
HD 124224&$855.5$ & $60140.55358$ & $60140.63711$ & $6.96 \pm 0.1$\\
HD 124224&$887.5$ & $59090.27544$ & $59090.28374$ & $2.5 \pm 0.02$\\
HD 124224&$887.5$ & $59417.43601$ & $59417.44441$ & $2.25 \pm 0.28$\\
HD 128898&$943.5$ & $59907.98747$ & $59908.40426$ & $0.44 \pm 0.03$\\
HD 133880&$943.5$ & $60337.93525$ & $60337.94573$ & $1.66 \pm 0.14$\\
HD 133880&$943.5$ & $60348.92369$ & $60348.93418$ & $5.18 \pm 0.34$\\
HD 133880&$1367.5$ & $59240.97039$ & $59240.98076$ & $2.21 \pm 0.14$\\
HD 133880&$1655.5$ & $59606.98603$ & $59606.9965$ & $1.87 \pm 0.2$\\
HD 134759&$943.5$ & $60337.97992$ & $60337.98971$ & $1.44 \pm 0.2$\\
HD 136347&$1367.5$ & $59240.99793$ & $59241.00841$ & $0.83 \pm 0.12$\\
HD 136504B&$943.5$ & $60338.91598$ & $60338.92646$ & $2.81 \pm 0.18$\\
HD 136504B&$943.5$ & $60348.92369$ & $60348.93418$ & $2.4 \pm 0.29$\\
HD 136504B&$1367.5$ & $59240.9841$ & $59240.99447$ & $4.59 \pm 0.7$\\
HD 136504B&$1367.5$ & $59275.82351$ & $59275.83398$ & $4.3 \pm 0.16$\\
HD 136504B&$1655.5$ & $59604.99038$ & $59605.00087$ & $2.44 \pm 0.2$\\
HD 136504B&$887.5$ & $58603.51118$ & $58603.77119$ & $2.38 \pm 0.07$\\
HD 136504B&$887.5$ & $58603.51118$ & $58603.77119$ & $3.08 \pm 0.02$\\
HD 142184&$842.5$ & $60125.59959$ & $60125.61366$ & $6.3 \pm 0.21$\\
HD 142184&$943.5$ & $60339.00825$ & $60339.01885$ & $5.01 \pm 0.15$\\
HD 142184&$887.5$ & $58598.53186$ & $58598.91975$ & $7.36 \pm 0.05$\\
HD 142184&$1367.5$ & $59232.03939$ & $59232.04986$ & $8.62 \pm 0.15$\\
HD 142184&$1655.5$ & $59599.04757$ & $59599.05817$ & $11.44 \pm 0.18$\\
HD 142184&$887.5$ & $58598.53186$ & $58598.91975$ & $7.36 \pm 0.05$\\
HD 142301&$842.5$ & $60125.59959$ & $60125.61366$ & $10.92 \pm 0.19$\\
HD 142301&$943.5$ & $60315.07267$ & $60315.08316$ & $2.91 \pm 0.32$\\
HD 142301&$887.5$ & $58598.53186$ & $58598.91975$ & $3.1 \pm 0.02$\\
HD 142301&$1367.5$ & $59232.03939$ & $59232.04986$ & $2.52 \pm 0.16$\\
HD 142301&$1655.5$ & $59595.08237$ & $59595.09297$ & $6.21 \pm 0.83$\\
HD 142301&$1655.5$ & $59599.04757$ & $59599.05817$ & $4.98 \pm 0.16$\\
HD 142301&$887.5$ & $58598.53186$ & $58598.91975$ & $3.1 \pm 0.02$\\
HD 145501C&$943.5$ & $60338.03605$ & $60338.04653$ & $1.16 \pm 0.1$\\
HD 145501C&$887.5$ & $58934.80448$ & $58934.81495$ & $3.78 \pm 0.01$\\
HD 145501C&$1367.5$ & $59274.94304$ & $59274.95353$ & $2.17 \pm 0.16$\\
HD 145501C&$1655.5$ & $59590.08179$ & $59590.09228$ & $1.75 \pm 0.2$\\
HD 145501C&$887.5$ & $58934.80448$ & $58934.81495$ & $3.78 \pm 0.01$\\
HD 147932&$1367.5$ & $60194.21015$ & $60194.54366$ & $13.45 \pm 0.04$\\
HD 147932&$943.5$ & $60327.06598$ & $60327.07659$ & $16.93 \pm 0.15$\\
HD 147932&$1367.5$ & $59232.05309$ & $59232.06358$ & $12.02 \pm 0.15$\\
HD 147932&$1655.5$ & $59599.06013$ & $59599.0705$ & $12.18 \pm 0.2$\\
HD 147932&$887.5$ & $58598.53186$ & $58598.91975$ & $6.45 \pm 0.03$\\
HD 147933&$1367.5$ & $60194.21015$ & $60194.54366$ & $4.02 \pm 0.03$\\
HD 147933&$943.5$ & $60327.06598$ & $60327.07659$ & $3.31 \pm 0.14$\\
HD 147933&$887.5$ & $58598.53186$ & $58598.91975$ & $2.12 \pm 0.01$\\
HD 148898&$1367.5$ & $60194.21015$ & $60194.54366$ & $0.2 \pm 0.03$\\
HD 148937&$1655.5$ & $59607.04434$ & $59607.05483$ & $3.04 \pm 0.4$\\
HD 149764&$887.5$ & $59949.0248$ & $59949.03333$ & $1.69 \pm 0.25$\\
HD 149764&$887.5$ & $60267.23021$ & $60267.23862$ & $2.08 \pm 0.27$\\
HD 149764&$1655.5$ & $59592.08831$ & $59592.0988$ & $1.03 \pm 0.16$\\
HD 149764&$887.5$ & $60468.60581$ & $60468.6141$ & $3.16 \pm 0.36$\\
HD 149764&$887.5$ & $60434.76943$ & $60434.77704$ & $1.54 \pm 0.26$\\
HD 149764&$887.5$ & $60324.07542$ & $60324.08383$ & $1.5 \pm 0.24$\\
HD 149764&$943.5$ & $60317.10024$ & $60317.11061$ & $2.2 \pm 0.18$\\
HD 149764&$943.5$ & $60317.08907$ & $60317.09944$ & $3.6 \pm 0.54$\\
HD 149764&$887.5$ & $60306.14628$ & $60306.15469$ & $1.32 \pm 0.17$\\
HD 149764&$887.5$ & $60291.21049$ & $60291.21902$ & $3.72 \pm 0.2$\\
HD 149764&$887.5$ & $60341.95056$ & $60341.95897$ & $10.58 \pm 0.27$\\
HD 149764&$887.5$ & $60250.28906$ & $60250.29735$ & $1.51 \pm 0.31$\\
HD 149764&$887.5$ & $60085.66806$ & $60085.67647$ & $1.4 \pm 0.19$\\
HD 149764&$887.5$ & $60235.34218$ & $60235.35047$ & $1.17 \pm 0.13$\\
HD 149764&$887.5$ & $59964.98047$ & $59964.98889$ & $1.51 \pm 0.3$\\
HD 149764&$887.5$ & $60025.8211$ & $60025.82951$ & $1.84 \pm 0.19$\\
HD 149764&$887.5$ & $60055.7414$ & $60055.7497$ & $1.34 \pm 0.23$\\
HD 149764&$887.5$ & $60100.61638$ & $60100.62466$ & $3.06 \pm 0.24$\\
HD 149764&$887.5$ & $60107.62957$ & $60107.63797$ & $1.51 \pm 0.2$\\
HD 149764&$887.5$ & $60146.54925$ & $60146.55778$ & $1.32 \pm 0.15$\\
HD 149764&$887.5$ & $60172.50851$ & $60172.51703$ & $2.84 \pm 0.2$\\
HD 151965&$887.5$ & $59949.03442$ & $59949.04272$ & $2.46 \pm 0.43$\\
HD 151965&$887.5$ & $60306.14628$ & $60306.15469$ & $2.04 \pm 0.24$\\
HD 151965&$943.5$ & $59635.70426$ & $59635.93351$ & $1.34 \pm 0.09$\\
HD 151965&$943.5$ & $59477.12612$ & $59477.54303$ & $1.71 \pm 0.03$\\
HD 151965&$943.5$ & $59391.38072$ & $59391.79751$ & $1.4 \pm 0.04$\\
HD 151965&$1367.5$ & $59231.08791$ & $59231.09851$ & $1.39 \pm 0.15$\\
HD 151965&$887.5$ & $60419.80543$ & $60419.81395$ & $2.53 \pm 0.21$\\
HD 151965&$1655.5$ & $59606.06001$ & $59606.07049$ & $1.84 \pm 0.14$\\
HD 151965&$887.5$ & $59979.03403$ & $59979.04233$ & $1.78 \pm 0.2$\\
HD 151965&$943.5$ & $60317.11131$ & $60317.12179$ & $1.72 \pm 0.2$\\
HD 151965&$887.5$ & $60250.28906$ & $60250.29735$ & $1.96 \pm 0.36$\\
HD 151965&$887.5$ & $60221.27722$ & $60221.28552$ & $1.77 \pm 0.34$\\
HD 151965&$887.5$ & $60203.37016$ & $60203.37858$ & $1.78 \pm 0.34$\\
HD 151965&$887.5$ & $60107.62957$ & $60107.63797$ & $1.95 \pm 0.16$\\
HD 151965&$887.5$ & $60291.21049$ & $60291.21902$ & $2.38 \pm 0.41$\\
HD 151965&$887.5$ & $59964.98047$ & $59964.98889$ & $2.19 \pm 0.26$\\
HD 151965&$943.5$ & $60339.03314$ & $60339.04374$ & $1.54 \pm 0.25$\\
HD 164224&$887.5$ & $60233.4005$ & $60233.40902$ & $1.86 \pm 0.25$\\
HD 164224&$887.5$ & $60512.51904$ & $60512.52745$ & $3.02 \pm 0.35$\\
HD 164224&$1367.5$ & $59231.12938$ & $59231.13986$ & $1.58 \pm 0.37$\\
HD 170397&$887.5$ & $59965.05466$ & $59965.06296$ & $1.33 \pm 0.25$\\
HD 170397&$1367.5$ & $59232.13604$ & $59232.1464$ & $1.12 \pm 0.16$\\
HD 170397&$887.5$ & $60342.01741$ & $60342.02594$ & $1.45 \pm 0.22$\\
HD 170397&$887.5$ & $59999.82259$ & $60000.0728$ & $1.06 \pm 0.12$\\
HD 171247&$887.5$ & $60417.81133$ & $60417.81986$ & $4.2 \pm 0.23$\\
HD 182180&$943.5$ & $60300.26642$ & $60300.27691$ & $7.19 \pm 0.15$\\
HD 182180&$887.5$ & $58601.83397$ & $58602.0949$ & $6.24 \pm 0.05$\\
HD 182180&$1367.5$ & $59210.24369$ & $59210.25405$ & $7.93 \pm 0.2$\\
HD 182180&$1655.5$ & $59605.16986$ & $59605.18035$ & $10.35 \pm 0.19$\\
HD 182180&$887.5$ & $58795.13728$ & $58795.22472$ & $5.56 \pm 0.03$\\
HD 182180&$887.5$ & $58601.83397$ & $58602.0949$ & $6.24 \pm 0.05$\\
HD 221006&$943.5$ & $60218.46277$ & $60218.87956$ & $0.29 \pm 0.02$\\
\end{longtable}
\end{center}
\clearpage
\twocolumn

\newpage

\begin{table*}[hbt!]
\footnotesize
\begin{threeparttable}
\caption{References for the stellar and magnetic parameters of the stars detected by ASKAP (Table \ref{tab:radio_detections}).}
\label{tab:references}
\begin{tabular}{lll}
\toprule
\headrow Star & Stellar parameters & Magnetic parameters\\
\midrule 
HD 12447& \citet{sikora2019a,sikora2019b} & \citet{sikora2019b}\\
HD 19832& \citet{netopil2008,shultz2020}& \citet{borra1983}\\
HD 22470& \citet{netopil2008,shultz2020}& \citet{borra1983}\\
HD 23478& \citet{sikora2015,shultz2019b} & \citet{sikora2015}\\
HD 25267& \citet{netopil2017,borra1980}& \citet{borra1980}\\
HD 27309& \citet{sikora2019a,north1995} & \citet{sikora2019b}\\
HD 34736& \citet{semenko2014,semenko2024}& \citet{semenko2024}\\
HD 35298& \citet{shultz2019b,shultz2018}& \citet{yakunin2013}\\
HD 35502& \citet{sikora2016}& \citet{sikora2016}\\
HD 36485& \citet{leone2010}& \citet{bohlender1987}\\
HD 36526& \citet{shultz2019b}& \citet{romanyuk2017}\\
HD 37017& \citet{bolton1998,shultz2018}& \citet{borra1979} \\
HD 37479& \citet{hunger1989,townsend2010}&\citet{shultz2020} \\
HD 37808& \citet{bernhard2020,romanyuk2021,shultz2022}& \citet{romanyuk2021,shultz2022}\\
HD 40312& \citet{sikora2019a,sikora2019b}& \citet{kochukhov2019}\\
HD 56350& \citet{shen2023}& \citet{bagnulo2015}\\
HD 61556& \citet{shultz2015}& \citet{shultz2015}\\
HD 64740& \citet{bohlender1990,shultz2018}& \citet{borra1979} \\
HD 83625& \citet{shen2023}, Das et al. under review & \citet{bagnulo2015}\\
HD 92385& \citet{netopil2017,renson2001}& \citet{landstreet2007}\\
HD 101412& \citet{folsom2012,hubrig2011a}& \citet{hubrig2011b}\\
HD 105382& \citet{briquet2001}& \citet{alecian2011} \\
HD 112381& \citet{netopil2017}& \citet{bohlender1993}\\
HD 112413& \citet{moiseeva2019,sikora2019b}& \citet{silvester2014,sikora2019b,romanyuk2016,romanyuk2017,romanyuk2020}\\
HD 118913& \citet{shen2023}& \citet{bagnulo2015}\\
HD 121743& \citet{alecian2014,shultz2018}& \citet{shultz2018} \\
HD 122451& \citet{pigulski2016,shultz2018,shultz2019b}& \citet{shultz2018}\\
HD 124224& \citet{sikora2019a,pyper1998,sokolov2000}& \citet{kochukhov2014,sikora2019b}\\
HD 128898& \citet{sikora2019a,kurtz1994,hummerich2016}& \citet{sikora2019b,bagnulo2015}\\
HD 133880& \citet{kochukhov2017}& \citet{bohlender1993}\\
HD 134759& \citet{buysschaert2018}& \citet{buysschaert2018}\\
HD 136347& \citet{landstreet2007}& \citet{landstreet2007}\\
HD 136504B& \citet{shultz2019b}& \citet{shultz2015b} \\
HD 142184& \citet{grunhut2012b}& \citet{grunhut2012b}\\
HD 142301& \citet{shultz2020}& \citet{borra1979,shultz2020} \\
HD 145501C& \citet{netopil2017,shultz2020}& \citet{borra1983,shultz2020}\\
HD 147932& \citet{shultz2020,rebull2018}& \citet{shultz2022}\\
HD 147933& \citet{leto2020}& \citet{pillitteri2018}\\
HD 148898& \citet{sikora2019a,sikora2019b}& \citet{sikora2019b}\\
HD 148937& \citet{naze2008,wade2012}& \citet{wade2012}\\
HD 149764& \citet{shen2023,renson2001}& \citet{bagnulo2015}\\
HD 151965& \citet{netopil2017,renson2001,das2025} & \citet{bohlender1993}\\
HD 164224& \citet{buysschaert2018,paunzen2015}& \citet{buysschaert2018}\\
HD 170397& \citet{netopil2017,renson2001}& \citet{borra1980,mathys1991}\\
HD 171247& \citet{netopil2017,renson2001}& \citet{shultz2022}\\
HD 182180& \citet{shultz2022,shultz2019b,rivinius2013}& \citet{rivinius2010,oksala2010,romanyuk2017,romanyuk2016}\\
HD 221006& \citet{netopil2017,renson2001}& \citet{bohlender1993}\\
\bottomrule
\end{tabular}
\end{threeparttable}
\end{table*}

\end{document}